\documentclass[fleqn,usenatbib]{mnras}

\usepackage{newtxtext,newtxmath}

\usepackage[T1]{fontenc}
\usepackage{ae,aecompl}

\usepackage{graphicx}	
\usepackage{amsmath}	
\usepackage{amssymb}	

\title[A revised lens time delay for JVAS~B0218+357]{A revised lens time delay for JVAS~B0218+357 from a reanalysis of VLA monitoring data}

\author[A. D. Biggs et al.]{
  A.~D.~Biggs$^{1}$\thanks{E--mail: abiggs@eso.org}
  and I.~W.~A.~Browne$^{2}$
  \\
  $^{1}$European Southern Observatory, Karl Schwarzschild Stra{\ss}e 2, D-85748 Garching, Germany \\
  $^{2}$Jodrell Bank Centre for Astrophysics, Alan Turing Building, School of Physics \& Astronomy, The University of Manchester, \\ Oxford Road, Manchester M13 9PL, UK
}

\date{Accepted XXX. Received YYY; in original form ZZZ}

\pubyear{2018}

\begin{document}
\label{firstpage}
\pagerange{\pageref{firstpage}--\pageref{lastpage}}
\maketitle

\begin{abstract}
  We have reanalysed the 1996/1997 VLA monitoring data of the gravitational lens system JVAS~B0218+357 to produce improved total flux density and polarization variability curves at 15, 8.4 and 5~GHz. This has been done using improved calibration techniques, accurate subtraction of the emission from the Einstein ring and careful correction of various systematic effects, especially an offset in polarization position angle that is hour-angle dependent. The variations in total and polarized flux density give the best constraints and we determine a combined delay estimate of $11.3 \pm 0.2$~d (1~$\sigma$). This is consistent with the $\gamma$-ray value recently derived using the \textit{Fermi Gamma-ray Space Telescope} and thus we find no evidence for a positional shift between the radio and $\gamma$-ray emitting regions. Combined with the previously published lens model found using \textsc{LensClean}, the new delay gives a value for the Hubble constant of $H_0 = 72.9 \pm 2.6$~km\,s$^{-1}$\,Mpc$^{-1}$ (1~$\sigma$).
\end{abstract}




\section{Introduction}
\label{introduction}

B0218+357 \citep{patnaik93} was revealed to be a gravitational lens by the Jodrell Bank--VLA Astrometric survey \citep[JVAS --][]{patnaik92} a search for unresolved calibrator sources that also detected a total of six gravitational lens systems \citep{king99}. With a separation of only 334~mas between the two lensed images (A and B) B0218+357 remains the smallest gravitational lens system detected to date and also contains an Einstein ring of approximately the same diameter. This system is well-studied, predominantly due to its suitability for measuring $H_0$ -- both the lensing galaxy and lensed source redshifts are known \citep*{browne93,cohen03}, the presence of a complete Einstein ring gives ample modelling constraints \citep{wucknitz04b} and the position of the lens galaxy has been independently measured using a deep \textit{Hubble Space Telescope} image \citep{york05}.

The time delay between images A and B has been measured several times via monitoring with the Very Large Array (VLA) in its largest `A' configuration. The first attempt at monitoring yielded an estimate of 12~d with a 1-$\sigma$ error of 3~d \citep{corbett96} from variations in the percentage polarization at 15~GHz. A subsequent campaign with more frequent monitoring determined a new value with much lower uncertainties ($10.5 \pm 0.2$~d; 1~$\sigma$) from variations seen in total flux, polarization and polarization position angle at 15 and 8.4~GHz \citep[][henceforth B99]{biggs99}. At the same time as the B99 observations, a separate monitoring campaign was taking place (with even higher sampling) that measured $10.1^{+1.5}_{-1.6}$~d (at 95~per~cent confidence) from total flux density data only \citep[][henceforth C00]{cohen00}.

\begin{table*}
  \centering
  \caption{Summary of VLA monitoring campaigns of B0218+357. The number of epochs refers to observations that produced calibratable data of the lens -- some epochs were terminated early due to technical problems, but appear in the archive.}
  \label{tab:obs}
  \begin{tabular}{ccccc} \\ \hline
    Project code & Dates & Frequency bands (GHz) & Number of epochs & Publication \\ \hline
    AP243 & 1992 Oct 2 to 1993 Jan 19 & 15, 8.4 & 25 & \citet{corbett96} \\
    AB707 & 1994 Mar 1 to 1994 May 30 & 15, 8.4 & 19 & Unpublished \\
    AH593 & 1996 Oct 10 to 1997 Jan 14 & 15, 8.4 & 60 & \citet{cohen00} \\
    AB809 & 1996 Oct 12 to 1997 Jan 14 & 15, 8.4, 5 & 46 & \citet{biggs99} \\ \hline
  \end{tabular}
\end{table*}

B0218+357 is also a $\gamma$-ray source and recent observations with the \textit{Fermi Gamma-ray Space Telescope} detected strong flaring activity which, despite the images being spatially unresolved, allowed a time delay to be determined \citep{cheung14} to very high accuracy. The $\gamma$-ray value ($11.46 \pm 0.16$~d; 1~$\sigma$) is significantly longer than the published radio delays and as noted by both \citeauthor{cheung14} and \citet{barnacka16} this could imply an offset between the radio and $\gamma$-ray emitting regions in the source plane (50--70~pc).

Given the discrepancy between the reported radio and $\gamma$-ray delays, and the potentially interesting astrophysical consequences of such a difference, it is important to try and improve the error bars on the delays. There is a potential to do this using existing archival radio data, not all of which have been published. C00 did not analyse their polarization data, whilst \citeauthor{corbett96} and B99 did not present their 8.4- and 5-GHz data respectively. Secondly, more densely sampled variability curves are possible as the B99 and C00 data were observed during the same observing season. Finally, improvements in the data analysis might be possible, particularly in the way that the Einstein ring emission is taken into account when deriving the flux densities of the two lensed images.

In this paper, we present the results of a re-analysis of the 1996/1997 VLA monitoring data. B0218+357 has to date been the subject of four separate observing campaigns at the VLA (Table~\ref{tab:obs}) three of which have been published. Two of these (AP243 and AB707) consisted of relatively few epochs and thus we do not consider these as they provide poorer constraints on the time delay. The 1996/1997 observations consist of 106 epochs, equivalent to an average sampling rate of better than one epoch per day. Great care has been taken to ensure the calibration of the data takes account of various systematic effects relating to how the observations were conducted (see Section~\ref{sec:analysis} -- we believe our analysis has a wider relevance to calibration of VLA data in general, but those readers not very interested in the calibration of VLA data might prefer to skip this section) and allowed a more robust measurement of the radio time delay and $H_0$ for this system. In addition, the multi-frequency data have allowed an investigation of the various Faraday effects that arise in the ISM of the lensing galaxy.

\section{Observations}
\label{observations}

All observations were conducted during the VLA's `A' configuration during October 1996 to January 1997 and used the then-standard continuum setup of two subbands (also known as IFs) each with a bandwidth of 50~MHz and sensitive to right and left circular polarizations (RCP and LCP). The average frequencies of the three frequency bands that were used to monitor the source were 14.9399, 8.4649 and 4.8601~GHz. At 15~GHz, the angular resolution of $\sim$120~mas allows easy separation of the two compact images and the contaminating steep-spectrum ring emission is relatively weak (e.g.\ fig.~1 of B99). On the other hand, at 15~GHz the gain of the telescope is a strong function of its elevation and the data are more affected by pointing errors (either systematic or due to wind) and bad (i.e.\ wet) weather. The biggest problem at the lower frequencies is the combination of lower angular resolution and brighter ring, which together make it more difficult to accurately measure the emission from the lensed quasar cores. At 5~GHz, the angular resolution is in fact comparable to the separation between A and B ($\sim$350~mas).

The two campaigns differed in their observing and calibration strategies. AH593 used 3C~48 to set the flux density of an unresolved gain calibrator close to the lens (J0205+3212), whilst AB809 used the very bright ($>$20~Jy) and very nearly unresolved 3C~84 as the flux and gain calibrator with 3C~119 used to check the variability properties of 3C~84. 3C~119 is well suited to this task as it is bright ($>$1~Jy), compact ($\la 100$~mas) and dominated by extended emission \citep{nan91} and thus unlikely to vary on short timescales. At 8.4~GHz, for example, the VLBI core constitutes $\la$5~per~cent of the total flux density measured by the VLA \citep{nan99,mantovani10}.

The instrumental polarization calibrator for both campaigns was 3C~84, an unpolarized source for which only a short observation is required. The polarization position angle (PPA) calibrator for AH593 was 3C~48. Although a standard VLA PPA calibrator, at these frequencies in `A' configuration it is very resolved and the PPA varies across the source. 3C~119 is also an excellent PPA calibrator as none of the total polarization ($>$10~per~cent at 15~GHz) seems to arise from the potentially variable core \citep{nan99,mantovani10}.

One significant difference between AB809 and AH593 is that in the former the pointing of the telescopes was improved by executing a reference-pointing scan after slewing to a new source. The \textit{a priori} pointing accuracy of a VLA telescope can be in error by up to 1~arcmin which, with a primary beam at 15~GHz of only 2.8~arcmin, can lead to significant reduction of the received signal. With reference pointing, the error can be reduced to as little as 2-3~arcsec.

Another factor affecting the pointing accuracy of the telescopes is the source elevation. Above $\sim$80\degr\ the alt-azimuth antennas of the VLA have difficulty in tracking a source and thus the pointing accuracy deteriorates causing errors in both the flux-density and PPA calibration, the latter due to errors in the assumed parallactic angle. This is mainly relevant for B0218+357 and the calibrators J0205+3212 and 3C~48, all of which pass within a few degrees of the zenith (the latitude of the VLA is 34.25\degr). Unfortunately, these sources were observed very close to transit on a number of occasions and this must be recognized when analysing the data.

The time spent observing B0218+357 per epoch varied, AB809 spending significantly more time at 15~GHz ($\sim$21~min) than at 8.4 (3~min) and 5~GHz (1~min). The time per epoch was less consistent for AH593, but the median values were 4 and 5~min at 8.4 and 15~GHz respectively.

\section{Data analysis}
\label{sec:analysis}

\subsection{Flux-density calibration}

\begin{figure*}
\begin{center}
\includegraphics[scale=0.21]{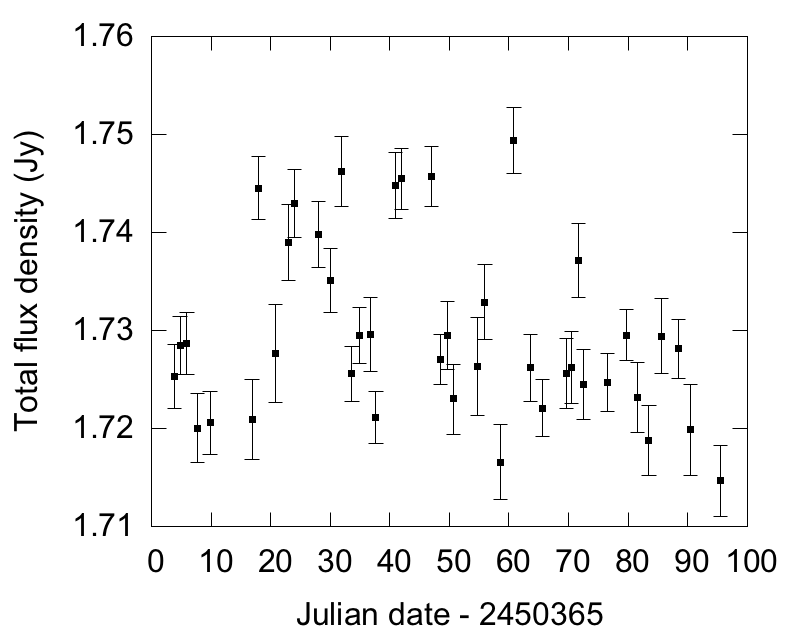}
\includegraphics[scale=0.21]{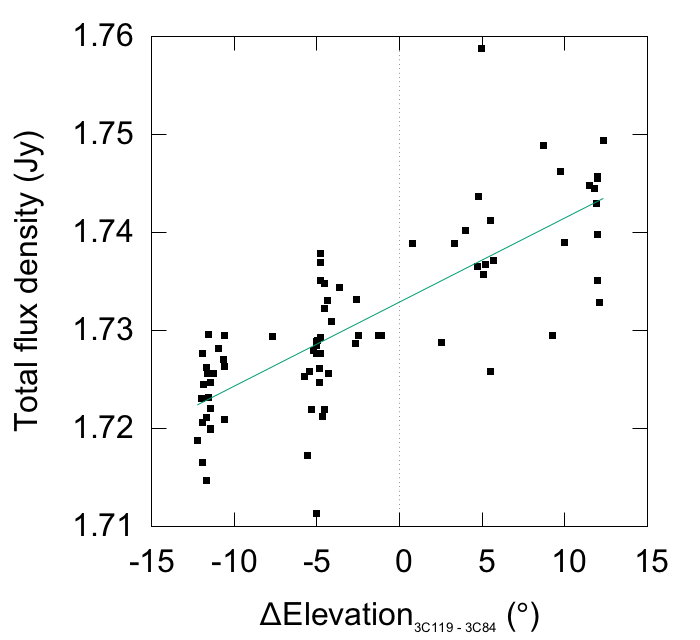}
\includegraphics[scale=0.21]{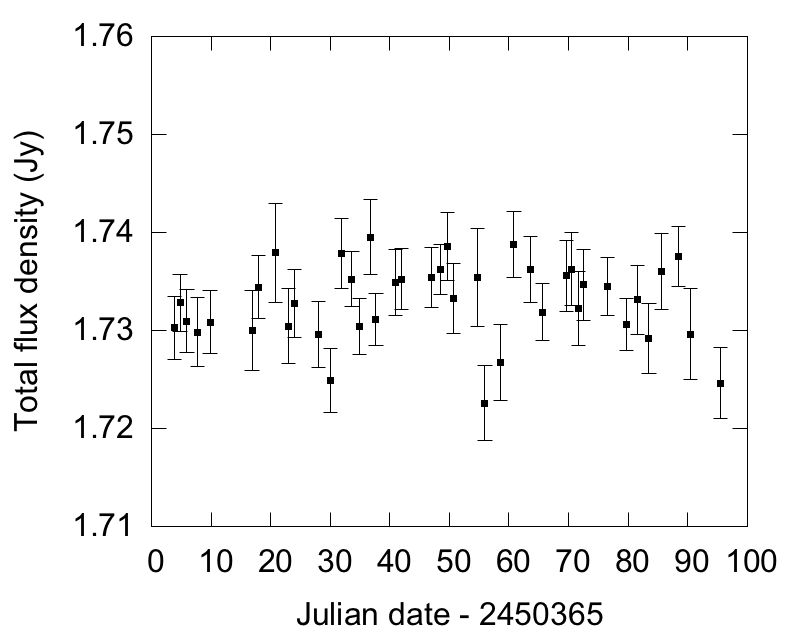}
\caption{Left: 15-GHz total flux density of 3C~119 calculated using the nearest 3C~84 scan in time. The higher flux density points correspond to epochs where 3C~119 is observed at higher elevation than 3C~84. Middle: 3C~119 flux density plotted as a function of the elevation difference between 3C~119 and 3C~84. As most epochs contain two 3C~84 scans typically separated by 30 minutes, calculating a 3C~119 flux using both allows the gain-elevation dependence to be better quantified. Also shown is a linear fit to the data. Error bars have been omitted for clarity. Right: the total flux density of 3C~119 corrected for the residual gain-elevation dependence. The rms scatter is $\sim$0.3~per~cent.}
\label{fig:3c119}
\end{center}
\end{figure*}

All data were calibrated using NRAO's Astronomical Image Processing System (\textsc{aips}) with the first step (common to all observations) being the application of \textit{a priori} curves that remove the elevation-dependence of the telescope gain. B99 found this step to be absolutely vital for detecting relatively small variations at 15~GHz that were nonetheless crucial in constraining the time delay. Although the gain-elevation dependence is much weaker at 8.4~GHz, we also apply the curves here.

For 3C~48, 3C~84 and 3C~119 we used clean-component source models during the calibration. For 3C~48 we used the models that come packaged with \textsc{aips}, whilst for the other two 3C sources we made our own source models by combining data from multiple epochs (for better $u,v$ coverage) and then producing a self-calibrated map.

For AH593, the calibration followed standard VLA procedures. The flux density of J0205+3212 was measured with respect to that of 3C~48, this having being calculated at each epoch using \textsc{setjy} which incorporates the time-dependence measured by \citet{perley13a}. As there were always two scans of J0205+3212 and often more than one of 3C~48, the flux density was estimated using the 3C~48 and J0205 scans that were closest in time.

The calibration of AB809 proceeded somewhat differently. The 3C~119 data were first calibrated assuming a constant flux for 3C~84. The resulting flux density of 3C~119 was seen to monotonically increase over the duration of the monitoring campaign, but this is a consequence of 3C~84 gradually fading since a large outburst many decades ago. We therefore measured the slope of the 3C~119 data and used this to set the 3C~84 flux density at each epoch such that the flux density of 3C~119 was constant with time. The calibration was then repeated. At both 15 and 8.4~GHz the flux density of 3C~84 monotonically declines by a few per~cent, consistent with measurements made with the single 26-m dish of the University of Michigan Radio Astronomy Observatory (UMRAO). An overall scaling factor was then computed by comparing the AB809 and AH593 0218+357 variability curves. As AH593 did not include observations at 5~GHz, we assumed that the flux density of 3C~119 at this frequency was that listed in the VLA Calibrator Manual, 3.7~Jy. This is consistent with the spectral index corresponding to the 3C~48-based measurements at 8.4 and 15~GHz.

\subsection{Residual gain-elevation correction}

The 15-GHz total flux density of 3C~119 is shown in Fig.~\ref{fig:3c119} where the error bars are those returned by \textsc{getjy}. The data are inconsistent with what would be expected for a non-variable source. Instead, the flux-density distribution is bimodal, approximately ten epochs being about 20~mJy brighter than the others. This is unlikely to reflect genuine variability in either 3C~84 or 3C~119 and in fact the key distinguishing feature of the higher flux-density epochs is the relative elevation of the calibrator and target.

This is demonstrated in the middle panel of Fig.~\ref{fig:3c119} where we plot the flux density of 3C~119 against the elevation difference between 3C~119 and 3C~84. In order to get better elevation coverage, we measure 3C~119's flux density at each epoch separately using both 3C~84 scans, these having been observed about 30~minutes apart. The flux density of 3C~119 appears to be a linear function of the elevation difference between it and its calibrator and is presumably a residual gain-elevation effect left after application of the \textit{a priori} curves.

The final panel of Fig.~\ref{fig:3c119} shows the 3C~119 15-GHz flux densities after correction for the residual gain-elevation effect. The scatter is significantly reduced (from $\sim$0.5 to $\sim$0.3~per~cent) and the bimodality has disappeared. This demonstrates not only the effectiveness of the correction, but also that 3C~84 is an excellent calibrator whose flux density varies linearly with time as we have assumed. A residual gain-elevation effect is also seen at 8.4~GHz which, as expected, is significantly weaker than at 15~GHz.
 
\begin{figure*}
\begin{center}
\includegraphics[scale=0.13]{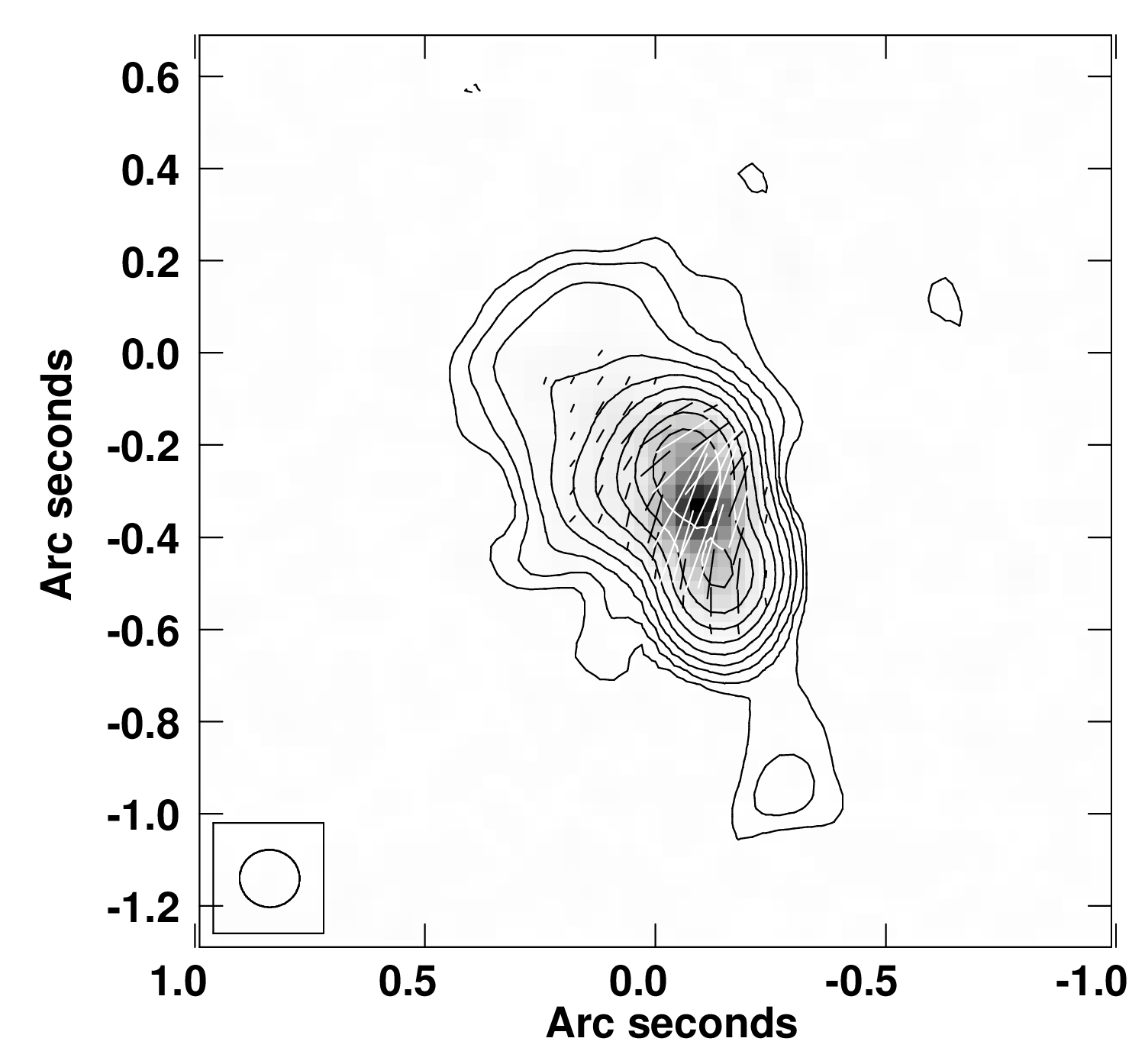}
\includegraphics[scale=0.13]{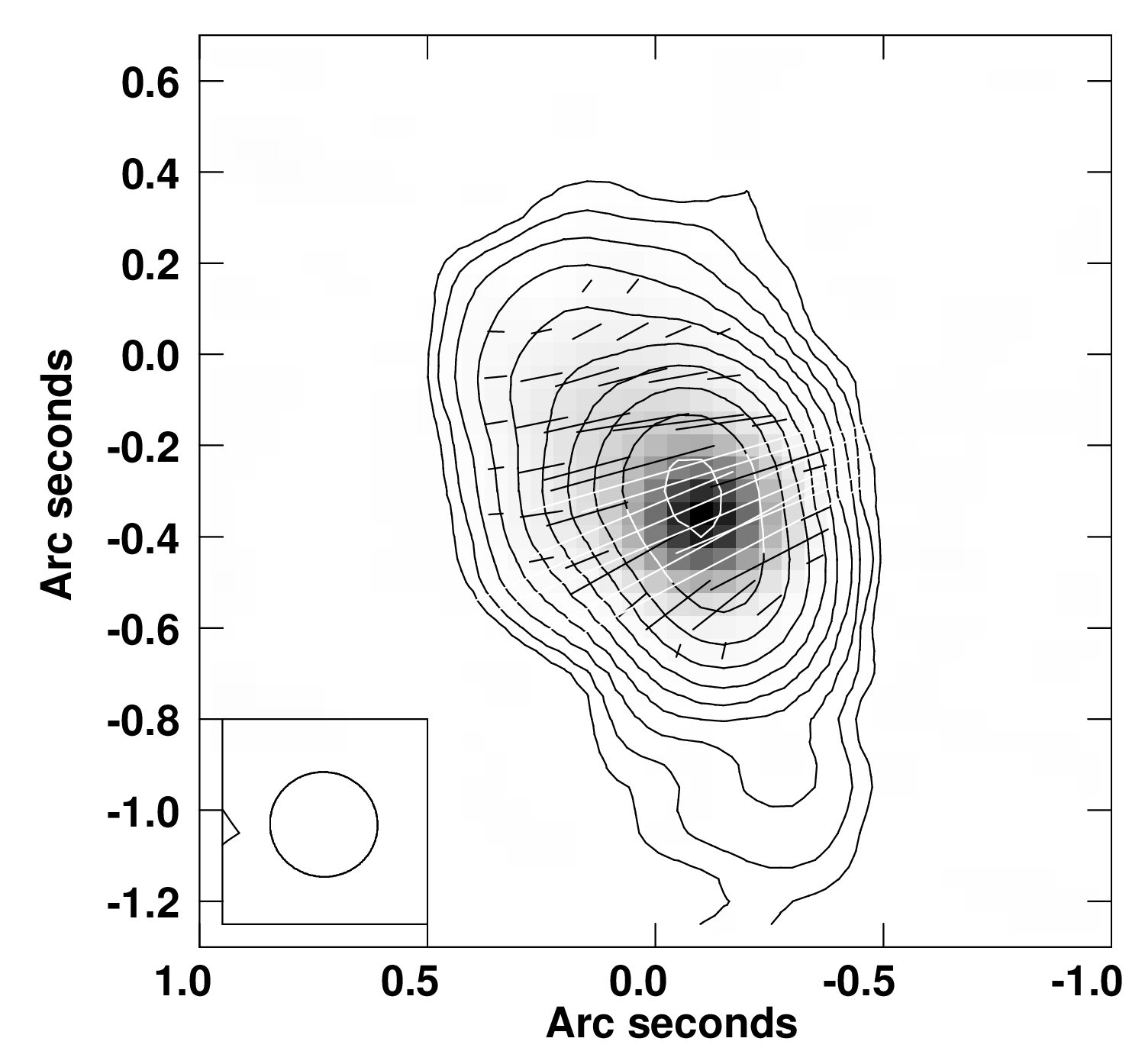}
\caption{Typical total-intensity maps of 3C48 (contours) at 15~GHz (left) and 8.4~GHz (right). These were made from the IF2 data of the 11 November epoch. The greyscale shows polarized flux, the peak pixel of which was used as the reference pixel for calibration of the PPA. Sticks show the orientation and magnitude of the polarization -- the peak polarized flux densities are 63 and 118~mJy\,beam$^{-1}$ at 15 and 8.4~GHz respectively. The synthesized beams are shown in the bottom left-hand corner. These are very nearly circular with a full width at half maximum of $\sim$130 and 230~mas$^2$.}
\label{fig:3c48map}
\end{center}
\end{figure*}

The residual gain-elevation effects should also be present in the 0218+357 variability curves and, as the elevation difference between 3C~84 and 0218+357 is similar to that between 3C~84 and 3C~119, should have a similar magnitude. However, due to the weaker flux densities of the lensed images the effect is weaker relative to the thermal errors and has no impact on the measured time delays. Despite this, the AB809 flux-density data presented in Section~\ref{sec:results} do have the correction applied. We have tried applying the same correction to the AH593 0218+357 data, but in this case the measured dispersion increases significantly and thus we use the uncorrected data in the time-delay analysis. The reason for the lack of improvement in this case is unknown, but may be related to the different calibration strategy whereby the flux scale is transferred from 3C~48 to 0205+322 and subsequently to 0218+357. 

\subsection{Polarization calibration}

For all data, the instrumental polarization (also known as polarization leakage or D-terms) was calculated using the task \textsc{pcal}, assuming that 3C~84 is unpolarized. With this removed from the data it was then possible to calibrate the PPA. The method used depended on whether 3C~119 or 3C~48 was the calibrator.

Although 3C~119 is moderately resolved in total intensity, the polarization structure at both frequencies is unresolved and therefore the $Q$ and $U$ flux densities were calculated by modelfitting a single delta component to the $u,v$ visibilities in \textsc{difmap} \citep{shepherd97}, for each IF separately. For simple source structures, modelfitting is preferable to mapping as there is no need to convolve/deconvolve and Fourier transform the data. The $Q$ and $U$ flux densities give the apparent PPA, which in turn gives the rotation in each IF necessary to produce the known PPA of 3C~119. This was measured with reference to 3C~286 using archival VLA data from 1989 (project AD238) that were taken in `D' configuration.

For 3C~48, modelfitting to the $u,v$ data is not possible as the polarization structure is significantly resolved and it was therefore necessary to make maps of each epoch's data. In order to do this as uniformly as possible, each Stokes $Q$ and $U$ map was made using an automated routine that made an initial map using 1000 clean iterations, measured the rms noise and then repeated the mapping such that this terminated once the rms noise was reached. We have used a single pixel (at the peak of the polarized-flux map) to act as a reference for the PPA i.e.\ each epoch's PPA was rotated such that the reference pixel gave the same value. Representative maps are shown in Fig.~\ref{fig:3c48map}.

The polarization variability curves of B0218+357, particularly polarization position angle, were found to be corrupted by a number of systematic errors that seriously degraded the quality of the variability curves. As both images of B0218+357 are bright and highly polarized, the thermal errors are relatively small and the fact that we have two copies of the intrinsic variability and many epochs greatly aids in identifying such effects that would otherwise be missed. The three effects that we found necessary to deal with were
\begin{enumerate}
\item poor reference antennas
\item an hour-angle-dependent PPA offset and
\item systematic errors in the polarization-leakage calibration.
\end{enumerate}

\subsubsection{Reference antenna}

As far as possible, we used the same reference antenna for all epochs, mostly for convenience when running the calibration scripts. However, whichever reference antenna is chosen, it is important for polarization calibration that its $\mathrm{RCP} - \mathrm{LCP}$ phase difference is stable with time as any variations will be transferred to all antennas during gain calibration and the measured source PPA will most likely be wrong. A reference antenna is usually selected that is close to the centre of the array as this reduces errors when phase referencing, but as we are actually self-calibrating all our sources in phase, any antenna could in principle be used. After investigating all possible reference antennas, antenna 25 (located on the W32 pad) was selected as the one which gave the best results.

\subsubsection{Hour-angle-dependent PPA offset}
\label{sec:hour_angle}

As many of the AH593 and AB809 epochs are observed close together in time, it is interesting to compare the $\mathrm{RCP} - \mathrm{LCP}$ phase rotation ($\Delta\phi_{RL}$) derived from the respective calibrators to see if they are in agreement. For each epoch of 3C48 we find the nearest 3C~119 scan, subtract the two phase corrections ($\Delta\phi_{\mathrm{RL,119}} - \Delta\phi_{\mathrm{RL,48}}$) and divide by two to convert to an error in PPA. As $\Delta\phi_{RL}$ is inherently variable, we ignore values where the time difference between the 3C~119 and 3C~48 scans is greater than 8 hours.

\begin{figure*}
\begin{center}
\includegraphics[scale=0.27]{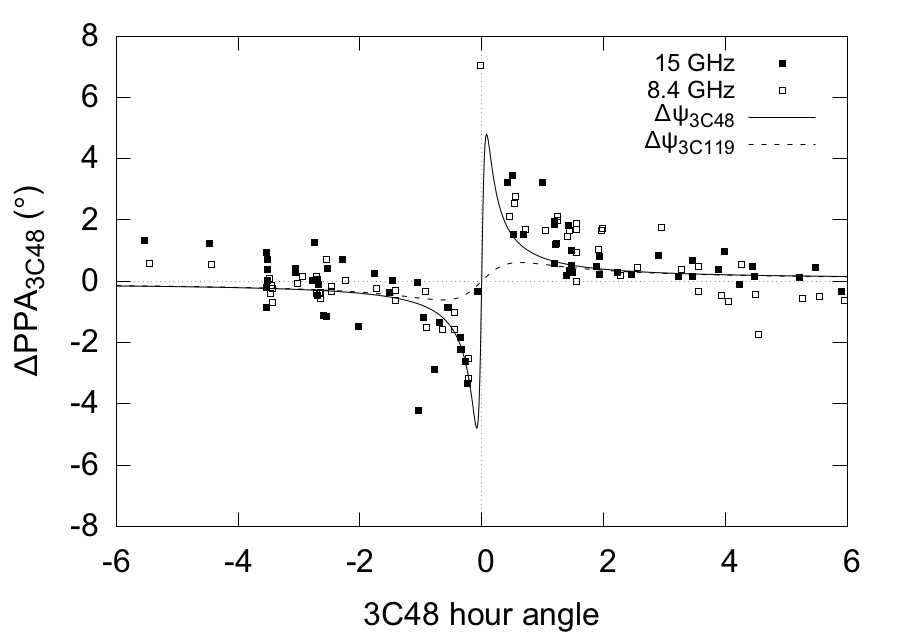}
\includegraphics[scale=0.27]{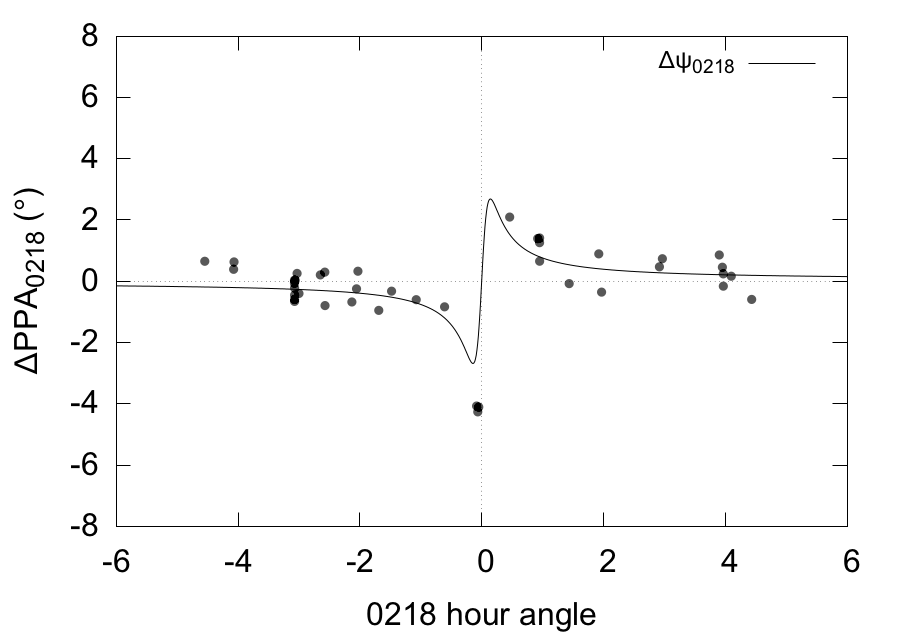}
\caption{Left: Error in the measured PPA of 3C48 derived from a comparison of the $\mathrm{RCP} - \mathrm{LCP}$ values between this source and 3C~119, assuming the latter to be correct. The data are plotted as a function of the 3C~48 hour angle and show a) that the error is negligible except around transit and b) that its sign reverses either side of this. Also plotted is a function that assumes that the error is due to an error in the parallactic angle, $\Delta \psi_{48}$. The much smaller error for 3C~119 is shown for comparison. Right: Residuals of the PPA of image B of 0218+357 at 8.4~GHz (AB809) after subtracting a polynomial fit to the variability curve. The same effect is clearly visible and is well described by the parallactic-angle error corresponding to the declination of 0218+357.}
\label{fig:nearest119}
\end{center}
\end{figure*}

The resultant values of $\Delta \mathrm{PPA}$ are plotted in the left panel of Fig.~\ref{fig:nearest119} as a function of 3C48's hour angle. The data show that the 3C~48 PPA calibration agrees well with that derived from 3C~119 over most of the hour-angle range, but is discrepant near, and reverses sign either side of, transit. Some form of systematic error is therefore affecting measurements of 3C48's PPA (the pattern disappears if the data are instead plotted against 3C~119's hour angle) and this seems to be the same at 8.4 and 15~GHz.

The effect is most likely related to the proximity of 3C48 to the zenith when observed close to transit and in particular to errors in the measurement of the parallactic angle, $\psi$. This measures the apparent rotation of a source (and hence its apparent PPA) with respect to the antenna feeds as a function of hour angle and must be removed if the correct PPA is to be measured. This is trivially calculated from knowledge of the antenna and source positions, but the observed trend corresponds to the functional form expected for an error in the parallactic angle due to the misalignment of an antenna pad (R.~Perley, private communication). The VLA was designed such that all antenna pads lie parallel to each other as this makes the parallactic angle the same for each. However, if a pad is tilted with respect to the others its parallactic angle will change as if the antenna were located at a different latitude.

As the offset in parallactic angle causes an equal rotation in PPA for a short observation, we calculate the error in the PPA as
\begin{equation}
\Delta \psi (\delta,H) = \psi(\delta,H,l_0) - \psi(\delta,H,l_0+\Delta l)
\end{equation}
where $\delta$ is the source declination, $H$ the source hour angle, $l_0$ the latitude of the VLA and $\Delta l$ the shift in apparent latitude caused by the misaligned pad. We find that $\Delta l = 10$~arcmin results in a good fit to the data ($\Delta \psi_{48}$ is also shown in Fig.~\ref{fig:nearest119}) although we note that this is much larger than the expected offsets of $1 - 2$~arcmin per pad (K.~Sowinski, private communication). A small number of antenna pads were erroneously aligned to local gravity (i.e. tilted away from the array centre) which could result in an error of up to 12~arcmin for a pad located at the end of an arm, but deleting each antenna in turn when mapping 3C48 does not remove the effect.

The same effect should also be present in the 0218+357 data as this source also passes close to the zenith. As the PPA variations at 8.4~GHz are relatively smooth they can be reasonably approximated using a third-order polynomial. The right panel of Fig.~\ref{fig:nearest119} shows the image-B residuals after subtracting both the polynomial fit and the contribution to the HA-based offset from 3C~119 i.e.\ $\Delta\psi_{\mathrm{119}}$. This should leave the PPA offset due to the elevation and hour angle of 0218+357. The theoretical variation of this, $\Delta\psi_{\mathrm{0218}}$, is also shown on the figure and fits the data quite well.

As the same effect is present in multiple sources and scales as expected with the source declination, we are confident that the $\Delta\psi$ curves describe the hour-angle-based PPA error very well and can be used to correct the variability curves.

\subsubsection{Systematic D-term offsets near transit}
\label{sec:dterms}

\begin{figure}
\begin{center}
\includegraphics[scale=0.27]{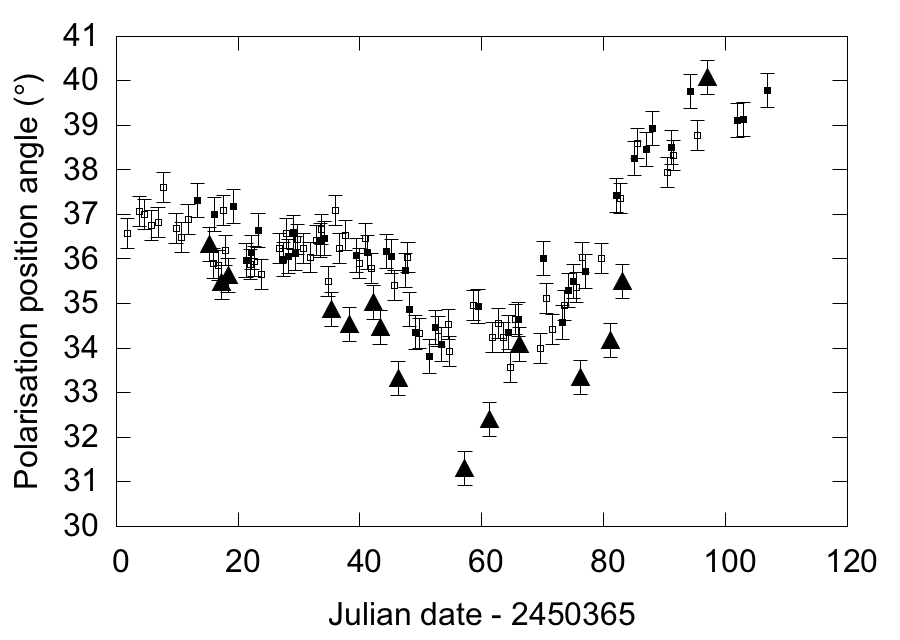}
\caption{PPA of both images of 0218+357 at 8.4~GHz as measured by the AH593 campaign -- the time delay and Faraday rotation have been removed. Image A is highlighted (large triangles) for those epochs where the 3C84 scan was observed within 1.2 hours of transit and nearly all can be seen to be offset by a few degrees in the same direction. This plot should be compared to that in Fig.~\ref{fig:abahlc_x} where it can be seen that the discrepant epochs have been successfully corrected.}
\label{fig:dterms}
\end{center}
\end{figure}

The final issue that we needed to correct for was only evident in the image-A PPA light curve of the AH593 data where approximately ten epochs had PPA measurements that were obviously too low by $1-3$\degr\ (Fig.~\ref{fig:dterms}). The factor that was common to each turned out to be the hour angle of the single scan of 3C84 that was used to calibrate the polarization leakage D-terms -- all were observed within approximately an hour of transit.

That the D-term calibration was the origin of the problem was confirmed by copying the D-terms from an AB809 epoch that was observed close in time and repeating the PPA calibration. In all cases this corrected the obvious problem with the PPA measurements and also led to improvements in the polarized-flux measurements where offsets had not been as evident. Therefore, all AH593 8.4-GHz epochs where 3C84 was observed within 1.2 hours of transit were calibrated in this way. However, in two cases where the AB809 3C~84 scans were observed either side of transit, a systematic PPA offset was visible in both image A and B of 0218+357. Copying the D-terms from the AB809 epoch next closest in time solved the problem.

\begin{figure*}
\begin{center}
\includegraphics[scale=0.11]{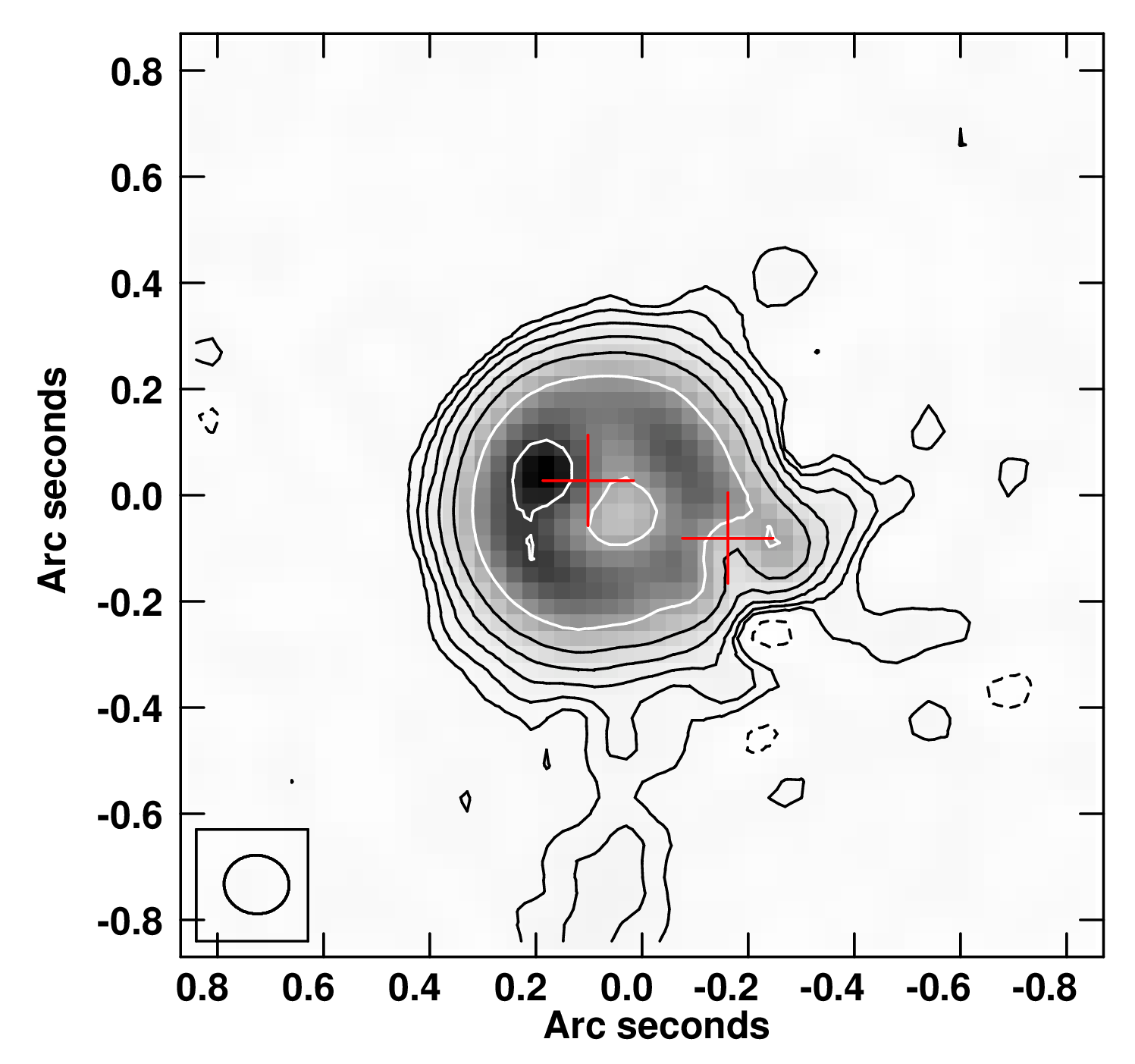}
\includegraphics[scale=0.11]{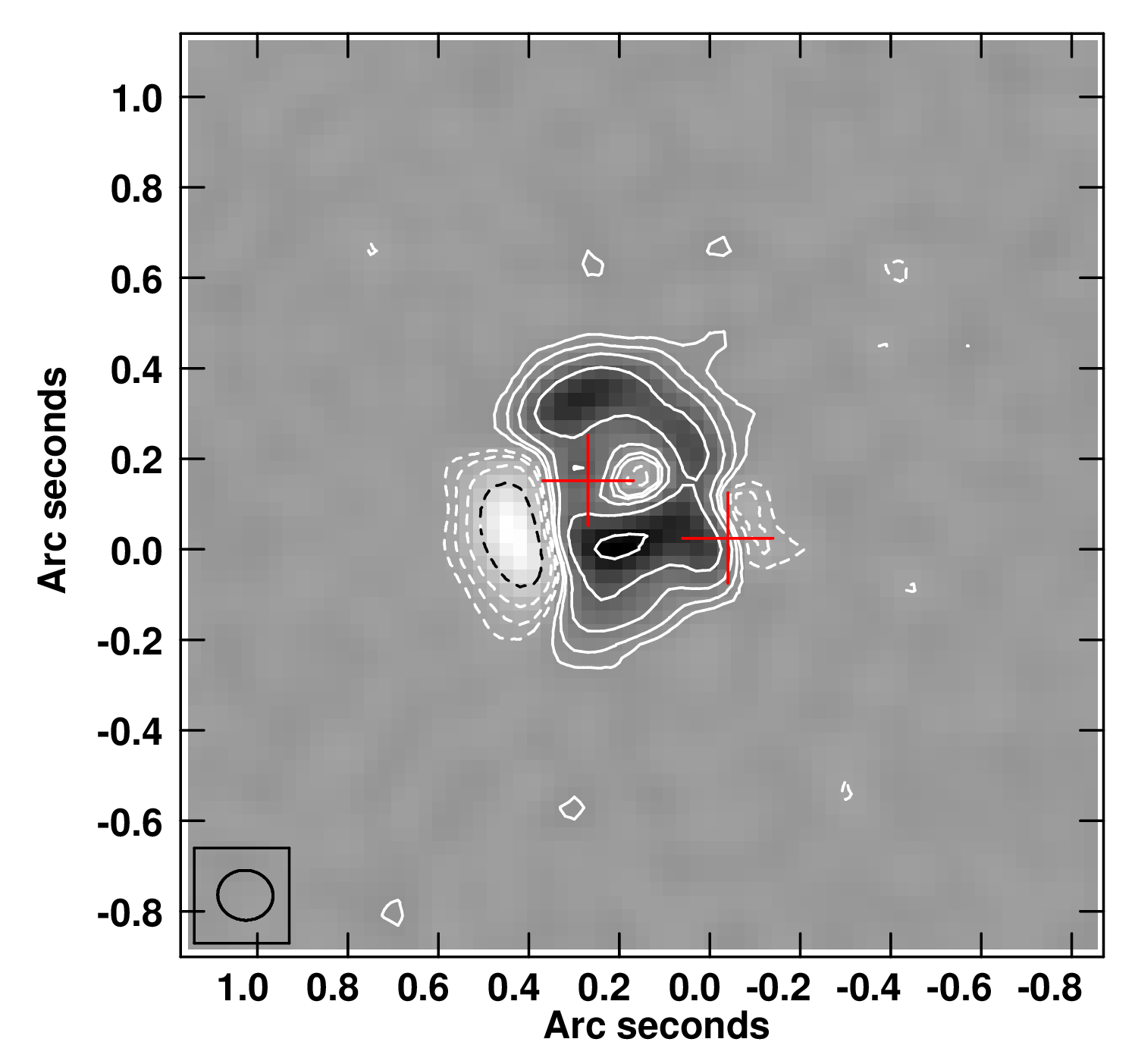}
\includegraphics[scale=0.11]{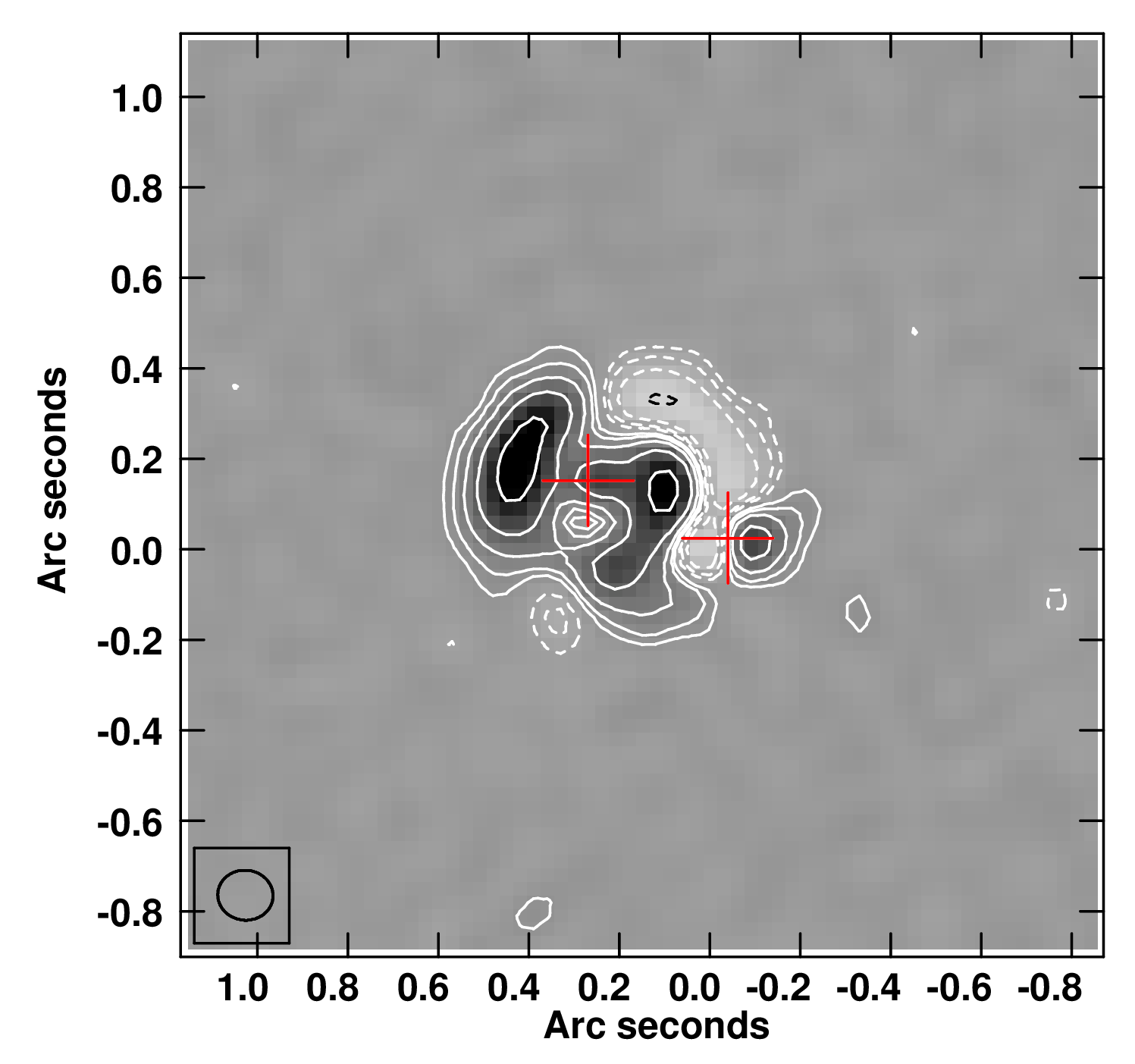}
\caption{Maps of the Einstein ring in (from left to right) Stokes $I$, $Q$ and $U$ at 15~GHz made from multiple epochs combined together. Each map's clean components were used to remove the ring emission from consideration when $u,v$ modelfitting. Similar models were also used at 8.4 and 5~GHz. The positions of A and B are marked with crosses.}
\label{fig:ringmod}
\end{center}
\end{figure*}

The reason why the D-term calibration is systematically in error only for these epochs of this campaign at this frequency is unknown. A possible clue comes from the fact that there are two things that distinguish the way 3C~84 was observed in the two campaigns: AB809 usually included two scans to AH593's one and had smaller pointing errors due to the use of reference pointing. The former results in more data and will tend to average out any hour-angle effects However, the third epoch had only one 3C~84 scan (due to problems with the reference pointing) and an hour angle of $-0.22$ hours. Despite this, no PPA offset is visible.

Problems with the pointing seem more likely given that this naturally deteriorates closer to transit for sources that pass this close to the zenith. This can potentially affect the polarization-leakage calibration as the instrumental polarization varies across the primary beam of the antenna \citep[e.g. fig~6.6 of][]{bignell82}. Whilst usually only considered significant for wide-field observations, it is possible that at the level of accuracy probed by our monitoring the effect is not negligible. However, it is difficult to understand how random pointing offsets could lead to such systematic offsets.

Why the problem is only seen in image A at 8.4~GHz is also not clearly understood. Our hypothesis is that the problems with the D-term calibration result in residual instrumental polarization that has a more-or-less constant position angle for each of the affected epochs. This will then combine vectorially with the source polarization which, at 8.4~GHz, is significantly different for each image due to Faraday rotation. Coincidentally, the PPA of the phase calibrator 0205+322 ($\sim$32\degr) is very close to that of image A at 8.4~GHz and the effect is also visible here.

\subsection{Modelfitting}

As with the 3C~119 PPA calibration, the flux densities of the lensed images A and B were measured by fitting model components to the calibrated $u,v$ data using \textsc{difmap}. A complicating factor is the Einstein ring, which if not dealt with would lead to poor fits and inaccurate flux densities. The approach followed by \citet{corbett96} was to simply exclude the short baselines ($<400$~k$\lambda$) as the jet emission responsible for the ring is resolved out on longer baselines. Unfortunately, for all 8.4-GHz epochs as well as those at 15~GHz observed at large hour angles, $>$50~per~cent of the data can be lost and the uncertainties in the flux densities increased. C00 instead mapped their data, whilst B99 modelled the ring with a Gaussian.

Our approach has been to produce the best model we can of the ring, in Stokes $I$, $Q$ and $U$. We do this by first fitting a model that only includes two delta components (A and B) to the Fourier-filtered data (baselines $>$400~k$\lambda$) of each epoch. The $I$, $Q$ and $U$ flux densities are then subtracted from the calibrated data in \textsc{aips} using the task \textsc{uvmod} and all epochs combined together. This dataset is then mapped to produce an image of the ring with excellent $u,v$ coverage and sensitivity (Fig.~\ref{fig:ringmod}). The clean components of the $I$ map are then subtracted from each epoch in \textsc{aips} and the two-component model fitted to all visibilities in \textsc{difmap}. As there is no easy way to subtract a multi-component $Q$ or $U$ model from a dataset in \textsc{aips}, these clean components are instead included in the \textsc{difmap} model.

As part of the modelfitting process we record the chi-squared ($\chi^2$) of the fit as well as the rms noise in each of the maps, $\sigma_I$, $\sigma_Q$ and $\sigma_U$.

\subsection{Error estimates and flagging of bad epochs}

The uncertainty on any measured value for each epoch has two components -- a random error due to thermal effects (time on-source, number of antennas, system temperature, weather, etc.) and an offset due to miscalibration of the flux scale or the $\mathrm{RCP} - \mathrm{LCP}$ phase difference (PPA).

The random errors have been calculated using the rms measured in each epoch's $I$, $Q$ and $U$ maps. Standard error propagation theory gives equations for the error on the polarized flux, $p$,
\begin{equation}
  \sigma_p = p \, \sqrt{Q^2 \sigma_Q^2 + U^2 \sigma_U^2}
\end{equation}
and the PPA,
\begin{equation}
  \sigma_{\mathrm{PPA}} = \frac{0.5}{1 + \left(\frac{U}{Q}\right)^2} \, \sqrt{\left(\frac{U \sigma_Q}{Q^2}\right)^2 + \left(\frac{\sigma_U}{Q}\right)^2}
\end{equation}
where the latter is given in radians.

The systematic flux-density offsets have been calculated in a statistical sense from the scatter in the 3C~119 flux densities. At each frequency we measure an rms of 0.3~per~cent, but conservatively assume 0.5~per~cent as the modelfitting errors for 0218+357 will likely be higher due to the more complex source structure. As the AH593 monitoring did not include a non-variable secondary calibrator it is not possible to derive a separate systematic uncertainty for this campaign and we therefore assume that the same value is also valid here. No systematic flux-calibration error affects the percentage-polarization measurements as these are formed from a flux-density ratio.

The systematic error on the PPA has been estimated by, for several epochs, calibrating with each possible reference antenna. The rms scatter of each epoch's PPA measurements lies between 0.2--0.4\degr and we adopt the larger value in order to be conservative. \citet{perley13b} have shown that 3C~48 is a very stable PPA calibrator at 8.4 and 15~GHz and the good agreement between the 3C~48 and 3C~119 PPA calibration demonstrates that the latter is also non-variable, as expected.

The final error on each measurement is derived by combining the random and systematic errors in quadrature. The errors on the percentage polarization are formed using the standard combination of the errors on the total and polarized flux densities.

Some epochs are removed completely from the final variability curves, for example if the weather was very bad (e.g.\ snow, thunderstorms) during the observations. This usually results in discrepant flux densities in both lensed images, poorer modelfits (as measured by chi-squared) and higher noise in a residual map. Observations very close to the zenith are often also corrupted as the antennas cannot track the source properly and three epochs have been flagged as 0218+357 was observed within 10~minutes of transit (antenna elevation $>$86\degr). In all three cases the image A and B total flux densities are obviously lower than the surrounding epochs and in two of three cases have significantly higher chi-squared values. In total we remove the same 14~epochs at 15 and 8.4~GHz. Five epochs are flagged at 5~GHz.

\section{Variability curves}
\label{sec:results}

\begin{figure*}
\begin{center}
\includegraphics[scale=0.42]{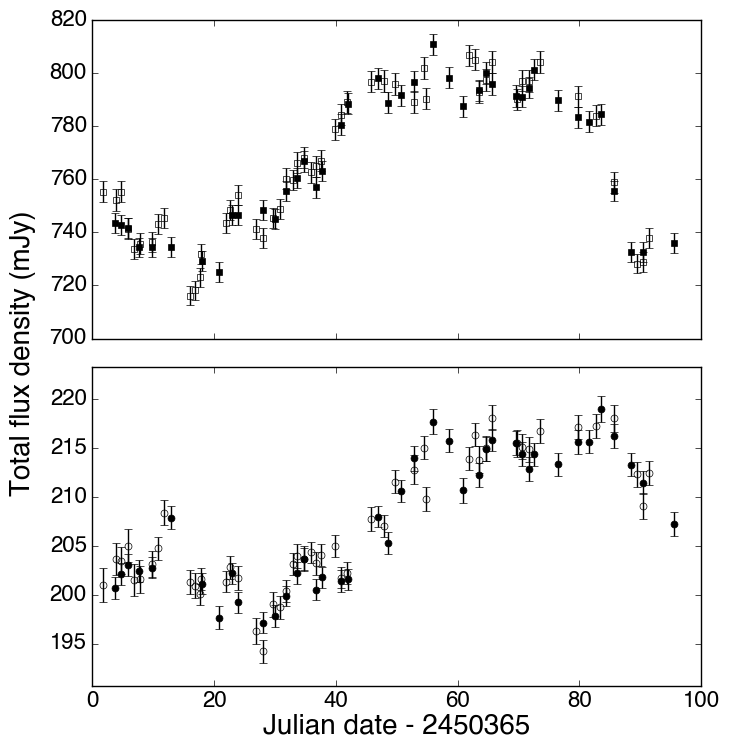}
\includegraphics[scale=0.42]{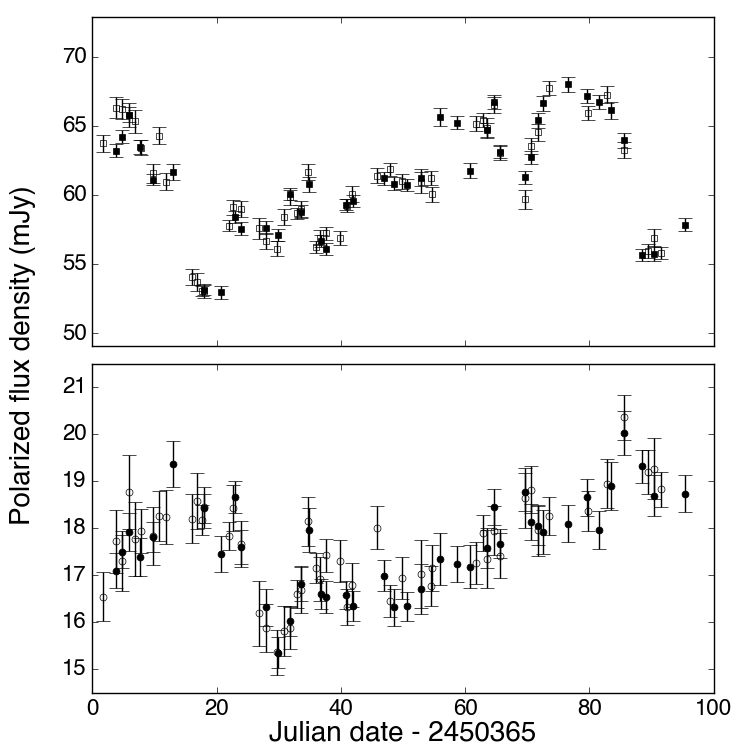}

\vspace{0.5cm}
\includegraphics[scale=0.42]{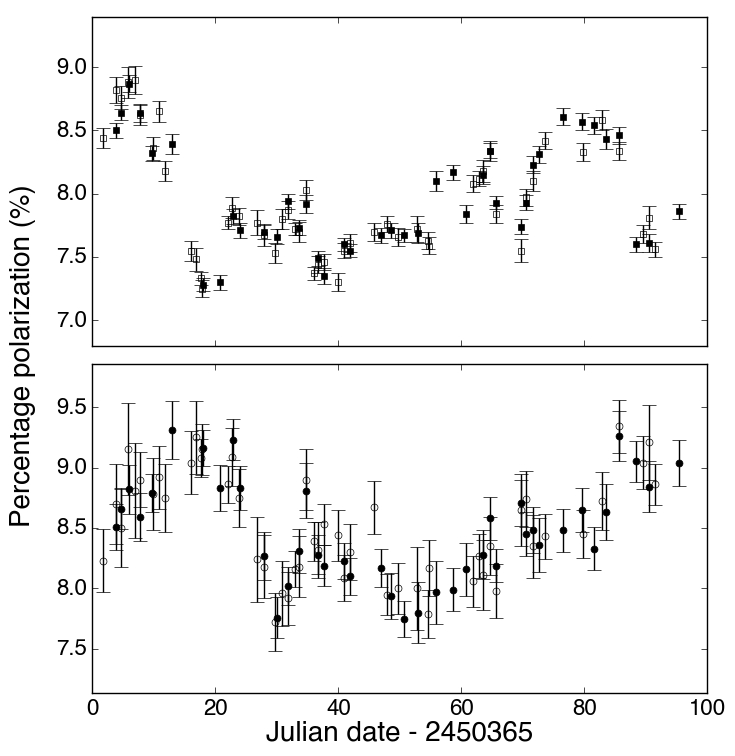}
\includegraphics[scale=0.42]{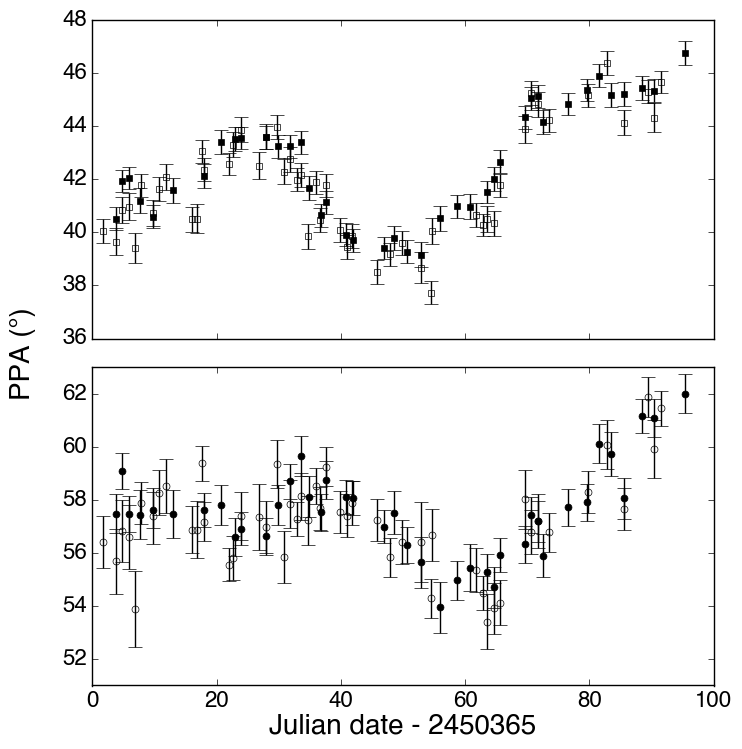}
\caption{Combined 15-GHz light curves of B0218+357 as determined from the AB809 (filled symbols) and AH593 (unfilled) monitoring campaigns. Top-left: total flux density, top-right: polarized flux density, bottom-left: percentage polarization and bottom-right: polarization position angle. Image A is the top panel in each case. The same $\Delta$PPA is used for A and B -- the other plots use the same fraction of the mean value.}
\label{fig:abahlc_u}
\end{center}
\end{figure*}

\begin{figure*}
\begin{center}
\includegraphics[scale=0.42]{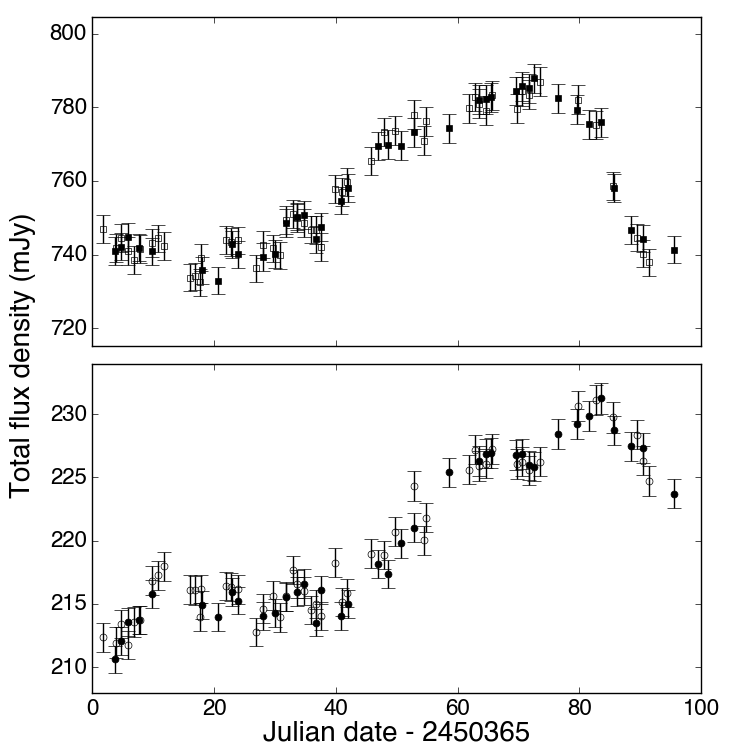}
\includegraphics[scale=0.42]{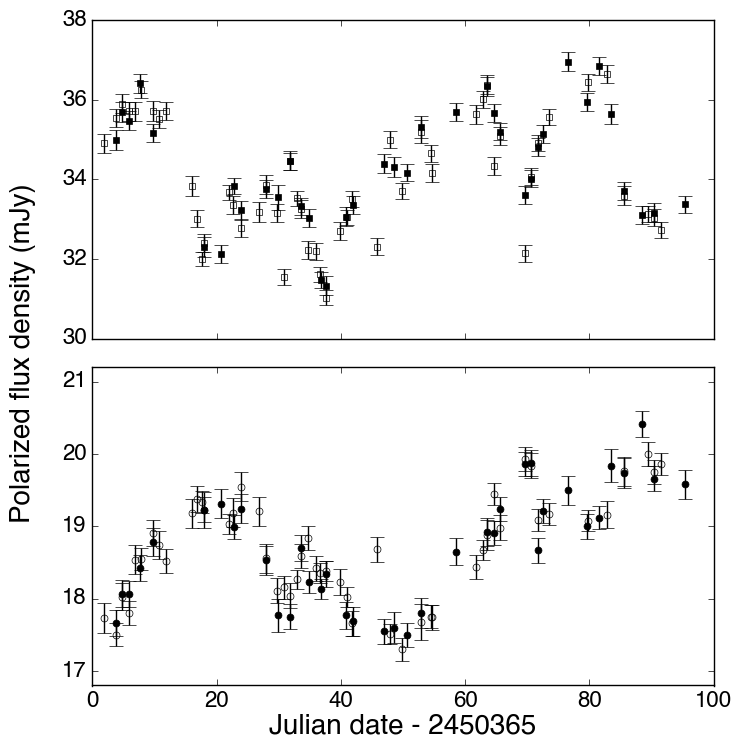}

\vspace{0.5cm}
\includegraphics[scale=0.42]{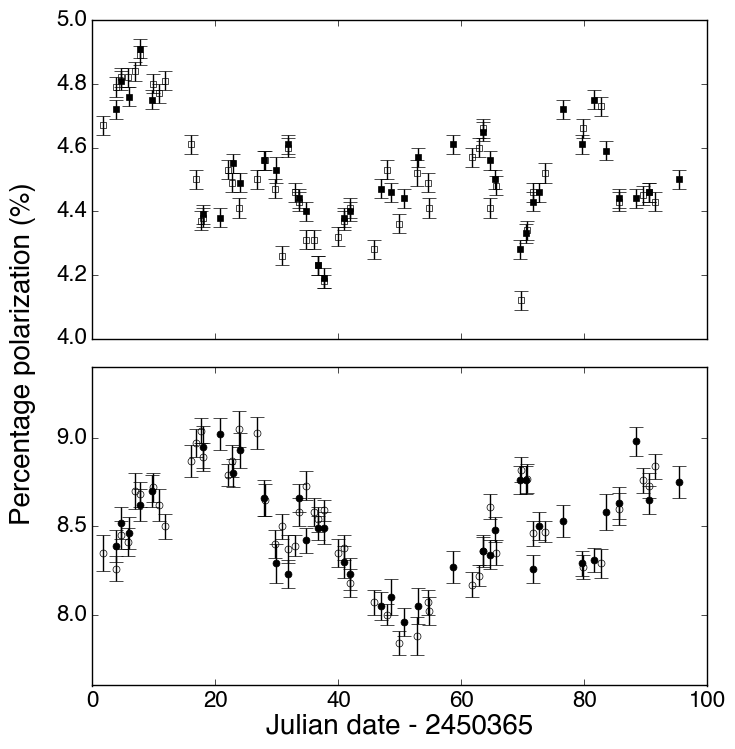}
\includegraphics[scale=0.42]{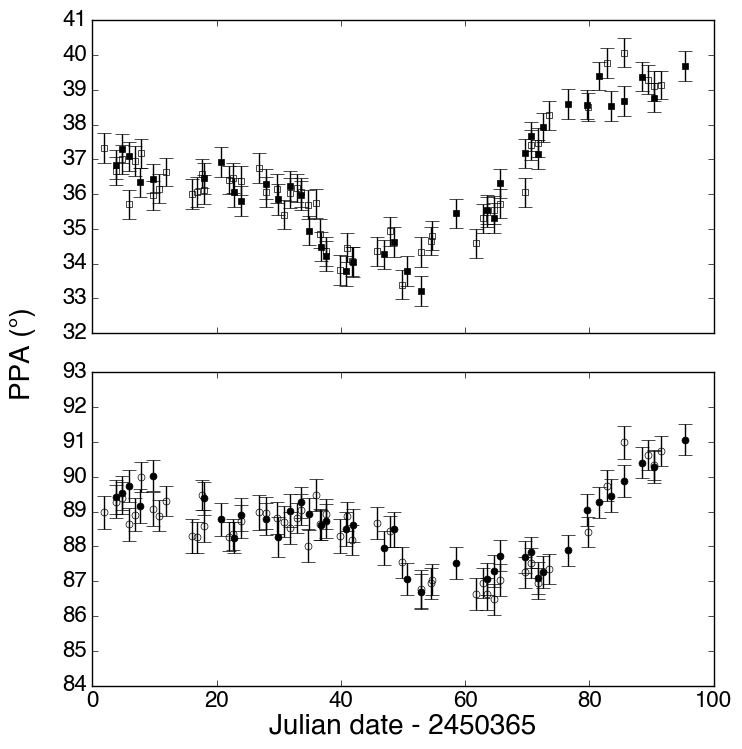}
\caption{Combined 8.4-GHz light curves of B0218+357 as determined from the AB809 (filled symbols) and AH593 (unfilled) monitoring campaigns. Top-left: total flux density, top-right: polarized flux density, bottom-left: percentage polarization and bottom-right: polarization position angle. Image A is the top panel in each case. The same $\Delta$PPA is used for A and B -- the other plots use the same fraction of the mean value.}
\label{fig:abahlc_x}
\end{center}
\end{figure*}

\begin{figure*}
\begin{center}
\includegraphics[scale=0.42]{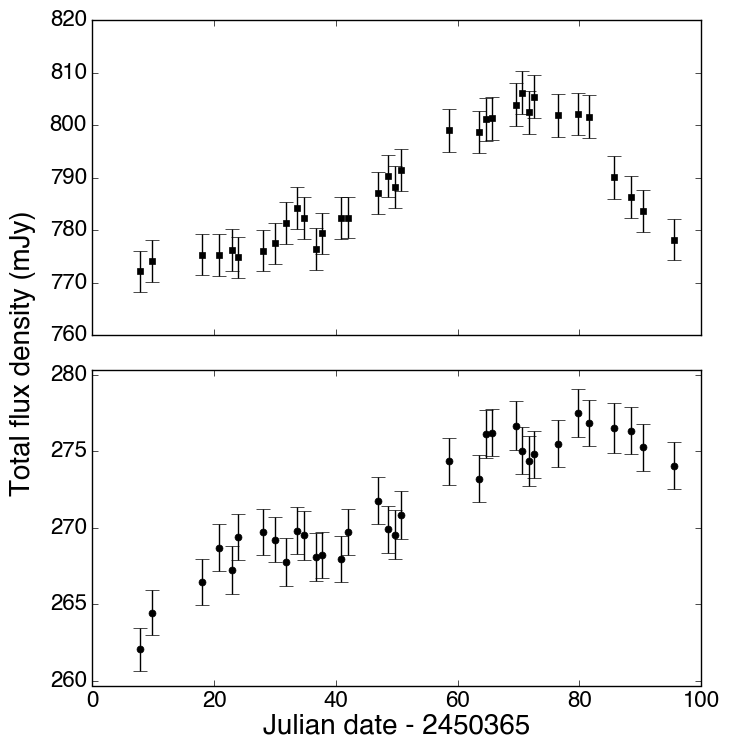}
\includegraphics[scale=0.42]{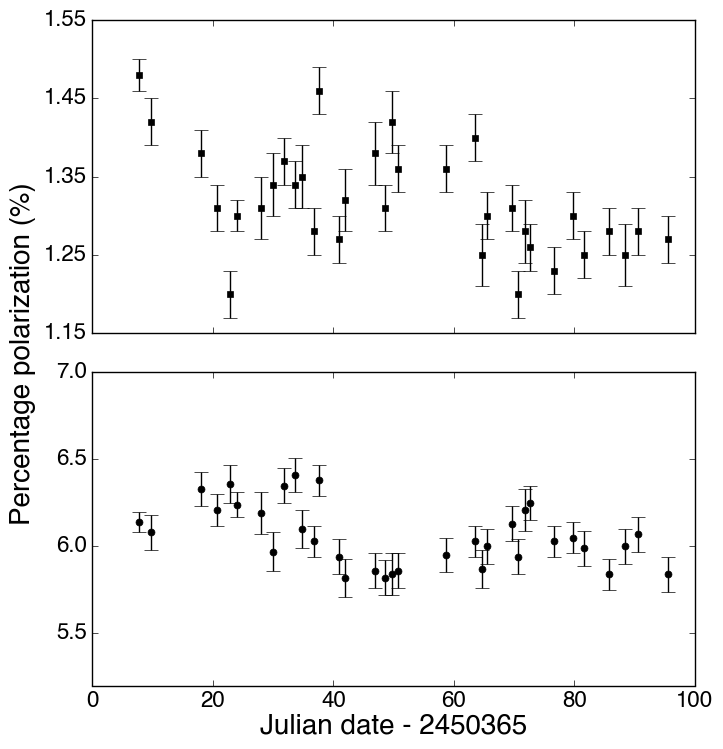}
\caption{5-GHz total flux density (left) and percentage-polarization (right) variability curves of B0218+357 as determined from the AB809 monitoring campaign. AH593 did not include any 5-GHz observations of 0218+357. Image A is the top panel in each case. Note the much more gradual decline of the flux density after day 80 and the low percentage polarization of image A compared to image B, indicating that A is highly depolarized. For each pair of A and B measurements, the $y$-range of each image is equal to the same fraction of the mean value.}
\label{fig:ab809lc_c}
\end{center}
\end{figure*}

The 0218+357 variability curves are shown in Figs~\ref{fig:abahlc_u} (15~GHz) and \ref{fig:abahlc_x} (8.4~GHz) and include plots of total flux density, polarized flux density, percentage polarization and the PPA. The 15-GHz and 8.4-GHz total flux density variations are very similar to those presented in B99 and C00, but much more densely sampled. Correlated variability is obvious in all plots with image B clearly lagging behind A.

Most improved compared to the previously published data are the 8.4-GHz polarization curves. B99 were unable to use these for measuring the time delay as the polarization measurements were extremely noisy, but our inclusion of the full $u,v$ range, careful ring subtraction and attention to systematic effects has reduced the thermal and systematic errors to such an extent that short-timescale variations common to both A and B are clearly visible. The variations seen at 8.4~GHz also broadly mirror those seen at 15~GHz, except for the expected differences due to depolarization (mostly in image A) and Faraday rotation.

Of the AB809 5-GHz data, we only show total flux density and percentage polarization (Fig.~\ref{fig:ab809lc_c}). The total-flux variations are similar to those seen at the higher frequencies and demonstrate that despite the relatively bright ring emission and the poor angular resolution, it is possible to measure the A and B total flux densities with reasonable accuracy. However, the magnitude of the variability is less and its timescale longer e.g.\ the much more gradual decline in image A towards the end of the monitoring.

The percentage-polarization data at 5~GHz demonstrate the much greater depolarization at this frequency, particularly in image A. Although the polarization of image B is reduced compared to 8.4~GHz, the reduction in image A is much more dramatic, this being a factor of $\sim$6 lower compared to 15~GHz. The variations seen at the higher frequencies are barely discernible in the 5-GHz data and it is clear that these cannot make a useful contribution to the time-delay analysis.

\section{Time-delay analysis}
\label{sec:time_delay}

\begin{figure}
  \begin{center}
    \includegraphics[scale=0.35]{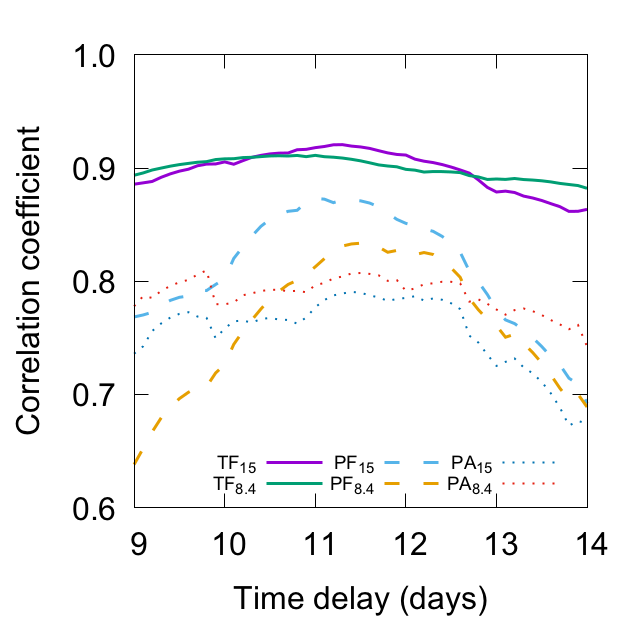}
    \caption{Correlation coefficient versus time delay for total flux (TF), polarized flux (PF) and PPA data. This shows that each dataset produces a global maximum around 11~d and also highlights the relative merits of each. The sharpest peak is seen for polarized flux density, whilst the CCF variation for PPA is less smooth and contains many pronounced local maxima.}
    \label{fig:ccf}
  \end{center}
\end{figure}

To measure the time delay we first begin by measuring the Pearson correlation coefficient, $r$, between A and B for various trial values of the time delay. For each delay, $\tau$, we shift A relative to B and linearly interpolate between the measured B values in order to produce a counterpart to each A measurement. The opposite is also done (B is shifted by the negative of the delay and image A interpolated) and the results averaged. This follows the procedure of \citet{white94}. Plots of $r$ versus time delay are shown in Fig.~\ref{fig:ccf} for each of the six datasets (total flux density, polarized flux density and PPA at 15 and 8.4~GHz) for delays between 0 and 20~d in steps of 0.1~d. The delays corresponding to the maximum correlation coefficient are given in Table~\ref{tab:delay}.

\begin{table}
  \centering
  \caption{Best-fit time delays for the cross-correlation (CCF) and chi-squared minimisation (CSM) techniques. As well as the six independent datasets, we also show the delays found using the flux-ratio data (Fig.~\ref{fig:fratio}). As these are not independent of the other measurements we do not include them in the final delay estimate.}
  \label{tab:delay}
  \begin{tabular}{ccccc} \\ \hline
    Dataset & Frequency (GHz) & \multicolumn{2}{c}{Time delay (d)} \\
     & & CCF & CSM \\ \hline
    Total flux density          & 15  & 11.3 & 11.3 \\
                                & 8.4 & 11.0 & 11.0 \\
    Polarized flux density      & 15  & 11.0 & 11.3 \\
                                & 8.4 & 11.5 & 11.5 \\
    Polarization position angle & 15  & 11.4 & 12.1 \\
                                & 8.4 &  9.6 & 12.8 \\ \hline
    Total flux density ratio    & 15 / 8.4 & 11.3 & 11.3 \\
    Polarized flux density ratio & 15 / 8.4 & 11.5 & 11.4 \\ \hline
  \end{tabular}
\end{table}

Fig.~\ref{fig:ccf} shows that the variation of $r$ with delay is similar for each frequency pairing e.g.\ total flux density at 8.4 and 15~GHz. This is as expected given that the variations are quite similar at each frequency. The highest correlations are seen for total flux density, whilst the maximum is most prominent (i.e.\ $r$ drops most rapidly away from the best delay) for polarized flux density, a consequence of the greater magnitude and shorter timescale of variability here. The PPA appears to offer the poorest constraints on the time delay as the $r$ curves are relatively shallow (particularly at 8.4~GHz) and contain many local maxima.

We have also performed a chi-squared minimisation (CSM). This uses the same interpolation scheme as the cross-correlation, but instead minimizes the square of the residuals (normalised by the square of the errors) summed over all overlapping points in the shifted variability curves. The algorithm also solves for the $y$-offset (e.g.\ flux ratio) between A and B. We note in passing that this is actually identical to the `dispersion minimization' presented by \citet*{tewes13}, as well as the `linear' technique used by \citet{kundic97}.

For five out of the six datasets we find $0.6 \le \chi^2 \le 1.2$ at minimum, thus confirming that the method for determining the error bars is reliable. For the 8.4-GHz polarized flux density the value is higher ($\chi^2 = 3.4$) which we attribute to interpolation errors caused by the rapid variability. The polarized-flux uncertainties are considerably smaller than at 15~GHz and this causes a proportionally larger increase in $\chi^2$ when gaps in the sampling cause the interpolation to become less accurate. As described below, this is compensated for when performing the final time-delay analysis.

The delays corresponding to the minimum (reduced) chi-squared are also shown in Table~\ref{tab:delay}. A comparison of the delays shows that the two algorithms give very similar results for total and polarized flux density at each frequency, the best-fit delays lying exclusively between 11.0 and 11.5~d. The PPA-based delays show a much wider spread, particularly at 8.4~GHz where the two values differ by $>$3~d, confirming that the PPA data are unable to strongly constrain the time delay. Excluding the PPA results, the reanalysed and combined radio data therefore favour a time delay that is much closer to the $\gamma$-ray value than that found by B99 and C00.

A fundamentally different approach to measuring the time delay is provided by the dispersion-minimization technique of \citet{pelt94,pelt96} as this does not use interpolation. For each trial delay and $y$-offset, this forms a combined variability curve and measures the agreement between nearby points. In the most commonly used variant of this statistic, $D^2_4$, pairs that are separated by $\la \delta$ (the so-called decorrelation length) are included, with a weight that decreases linearly with their separation. There is no simple answer as to what value to select for $\delta$ and in practice this parameter must be varied to investigate its effect on the best-fit time delay. Larger values include more pairs and reduce the noise in the variation of the dispersion with time delay, but can also bias the time delay from the true value.

\begin{figure}
  \begin{center}
    \includegraphics[scale=0.35]{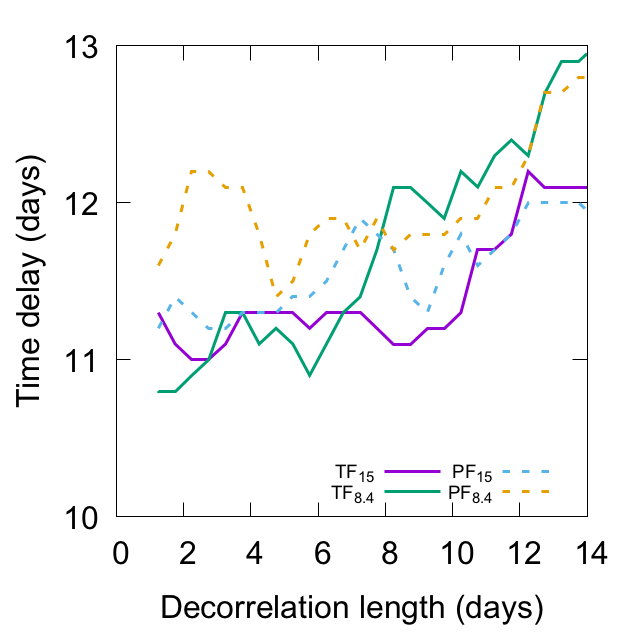}
    \caption{Best-fit time delay found using the Pelt dispersion method ($D^2_{4,2}$) as a function of the decorrelation length $\delta$ for the non-PPA datasets only. The increase in the time delay for large values of $\delta$ is probably due to the `curvature bias' described by \citet{pelt96}. For $\delta \la 6$~d the best-fit delays inhabit a very narrow range of approximately $11 \le \tau \le 12$~d.}
    \label{fig:delta}
  \end{center}
\end{figure}

In Fig.~\ref{fig:delta} we show the time delay at minimum dispersion for the four non-PPA datasets as a function of $\delta$ ($1 \la \delta \la 14$~d). There is a clear trend for the delay to increase for larger values of $\delta$ and we identify this with the `curvature bias' described by \citet{pelt96}. For $\delta \la 6$~d the delays are much more stable and three (total flux density at both frequencies and the 15-GHz polarized flux density) produce delays that are consistent with the range found using the cross-correlation and chi-squared analyses. The 15-GHz total flux density is particularly stable with $11.0 \le \tau \le 11.3$~d for $\delta \la 10$~d. In common with the CCF and CSM methods, the polarized flux density at 8.4~GHz gives slightly longer delays.

In order to investigate the uncertainties on the time delays in more detail, we have performed Monte Carlo analyses on the non-PPA datasets, using both model-independent and -dependent approaches.

Firstly, we have performed a bootstrap analysis using the cross-correlation algorithm following the FR/RSS method of \citet{peterson98}. This produces multiple realisations of a dataset by randomly sampling (with replacement) from the available epochs, perturbing each measurement within its error bar and performing the cross-correlation analysis on each. This produces a conservative estimate of the error on the time delay as duplicated epochs (produced due to sampling with replacement) are excluded i.e.\ only a subset of the epochs is used to measure the time delay each time. However, a strength of this technique is that it is model-independent i.e.\ no assumption is made about the intrinsic source variability.

The median delay and the bounds corresponding to the 16th and 84th percentiles ($\pm 1 \sigma$) are shown in Table~\ref{tab:bootstrap} and are the result of 5000 realisations of each dataset. The tightest constraints are provided by the total flux density at 15~GHz and the two polarized flux density datasets which have a total 1-$\sigma$ width of $\le 1$~d. The time-delay distributions found for the PPA datasets cover a very broad range and consist of multiple separated groups of delays. It is therefore impossible to assign a sigma to them and we do not include them in Table~\ref{tab:bootstrap} or consider them further.

\begin{table*}
  \centering
  \caption{Time delays and 1-$\sigma$ error bounds found using Monte Carlo techniques together with the cross-correlation (CCF) and chi-squared minimisation (CSM) techniques. The flux ratios are also shown for the $\chi^2$ method. The results for the PPA data are not shown due to their very poor constraints on the time delay. The flux density ratios differ significantly from those presented by B99 due to the different strategies used to remove the ring emission. The new values should be more accurate.}
  \label{tab:bootstrap}
  \begin{tabular}{cccccc} \\ \hline
    Dataset & Frequency (GHz) & \multicolumn{2}{c}{Time delay (d)} & Flux ratio \\
    & & CCF & CSM & CSM \\ \hline
    Total flux density          & 15  & $11.3^{+0.2}_{-0.7}$ & $11.3 \pm 0.4$ & $3.689 \pm 0.004$ \\
                                & 8.4 & $10.5^{+0.9}_{-1.3}$ & $11.0 \pm 0.7$ & $3.429 \pm 0.003$ \\
    Polarized flux density      & 15  & $11.4^{+0.4}_{-0.6}$ & $11.3 \pm 0.3$ & $3.470 \pm 0.015$ \\
                                & 8.4 & $11.6^{+0.6}_{-0.3}$ & $11.5 \pm 0.4$ & $1.825 \pm 0.005$ \\ \hline
  \end{tabular}
\end{table*}

As an alternative approach, we have used the best-fit time delay and $y$-offset from the CSM algorithm to form a combined variability curve and fitted a smoothing spline (using the \textsc{UnivariateSpline} function from \textsc{scipy}) to this. In doing so we conservatively scale the error bars such that the minimum $\chi^2 = 1$. This has the greatest effect for the 8.4-GHz polarized flux density and produces a much smoother fit than if the calculated uncertainties had been used. In all cases the fitted spline reproduces the observed variability well. Simulated variability curves are then produced by sampling the spline at new time intervals formed from randomly selecting from the original sampling intervals and then perturbing each point randomly within its error bar.

\begin{figure}
  \begin{center}
    \includegraphics[scale=0.35]{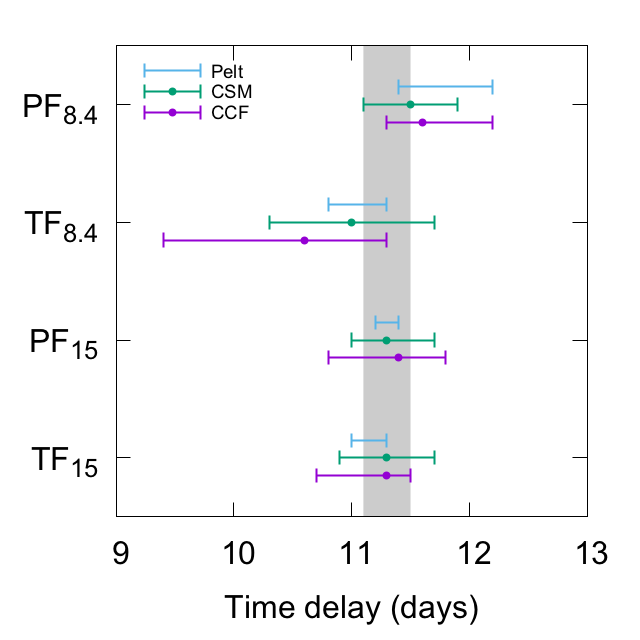}
    \caption{Time delays and uncertainties for the four non-PPA datasets measured using the cross-correlation (CCF) and chi-squared minimisation (CSM) techniques. Also shown is the range of delays found using the Pelt $D^2_{4,2}$ statistic for $\delta < 6$~d. The vertical grey band indicates the final revised time delay estimate of $11.3\pm0.2$~d (1~$\sigma$).}
    \label{fig:delaycomp}
  \end{center}
\end{figure}

The results are also shown in Table~\ref{tab:bootstrap}, together with the median and 1-$\sigma$ bounds for the $y$-offsets. The delay distributions are actually quite similar to those found with the cross-correlation bootstrapping, although in general a little narrower. The median delays and their 1-$\sigma$ errors for both techniques are shown in Fig.~\ref{fig:delaycomp} together with the range of delays found using the $D^2_{4,2}$ statistic for $\delta < 6$~d. For each dataset, the results found from the different techniques are consistent at 1~$\sigma$.

In order to combine the time delays found from the individual datasets into a final time-delay estimate, we have taken a weighted average of the results. This is a completely different (and much more straightforward) method to that used by B99 and is possible as the delay distributions are symmetric Gaussians. We use the results from the $\chi^2$ analysis as these should have more realistic error bars. Averaging assumes that each dataset should give the same delay and this assumption is based on the lack of a position shift ($<$1~mas) between 8.4 and 15~GHz \citep{mittal06}. This gives a final delay estimate of $\tau = 11.3 \pm 0.2$~d. This is plotted as a grey bar on Fig.~\ref{fig:delaycomp} where it can be seen that it is compatible with the delay range found for every dataset and time-delay algorithm combination. The weighted average of the values found from the CCF technique equals 11.4~d.

\section{Discussion}
\label{discussion}

\subsection{The revised radio time delay}

Our revised radio time delay for B0218+357 ($\tau = 11.3 \pm 0.2$~d; 1~$\sigma$) is higher than that previously published by B99 (10.5~d) and C00 (10.1~d) and statistically inconsistent with the B99 result, for which the 1-$\sigma$ error bar of 0.2~d seems to have been substantially underestimated. Looking at the B99 results in more detail, the individual datasets returned time delays that were consistent with the revised value except in the case of the 15-GHz PPA. We now know that the PPA measurements are subject to a systematic offset that is dependent upon the hour angle (Section~\ref{sec:hour_angle}) and this may have biased the analysis towards shorter delays. It is also the case that the dataset with the lowest uncertainty, the 15-GHz percentage polarization, was most consistent with the revised delay.

The reanalysed data are far superior to those presented by B99. Each dataset consists of $\sim$150~per~cent more epochs, the error bars have been more rigorously derived (especially for polarization) and various systematic effects have been removed by careful subtraction of the Einstein ring, removal of a residual gain-elevation effect, a better choice of reference antenna, correction for the dependence of the PPA on the hour angle and a more careful D-term calibration at 8.4~GHz. Our time-delay analysis is also more rigorous. We have presented the results from three different techniques, and applied two variants of Monte Carlo simulations to two of these. We find that {\em every} range of delays found using the techniques is consistent with our best-fit delay and error for every dataset.

Taken individually, the B99 results are consistent at 1~$\sigma$ with the new time delay except for PPA at 15~GHz. This gives us confidence that our basic technique for deriving time delays and their associated uncertainty is sound. The main reason for the offset in the B99 result is likely to have been the inclusion of the PPA dataset that we now know to be systematically biased, as well as the technique used to combine the individual time delays into a global average. B99 used a somewhat ad hoc `simultaneous chi-squared minimization', but our new analysis uses a much more standard and straightforward weighted average.

The PPA datasets have in general proven to be less useful for measuring the time delay to high precision. This is to a large extent due to the relatively slow variability timescale and, particularly at 8.4~GHz, the low magnitude of the variability. It remains possible though that systematic errors related to the HA-based offsets and D-term calibration are still present in the data. The discovery of several systematic effects related to the polarization calibration has been disappointing, particularly as polarization monitoring should in principle produce particularly robust variability curves as they do not depend on the absolute flux calibration. Other lens systems that do not pass as close to the zenith should be less affected, although we caution that we do not fully understand the origin of these effects.

\begin{figure*}
  \begin{center}
    \includegraphics[scale=0.27]{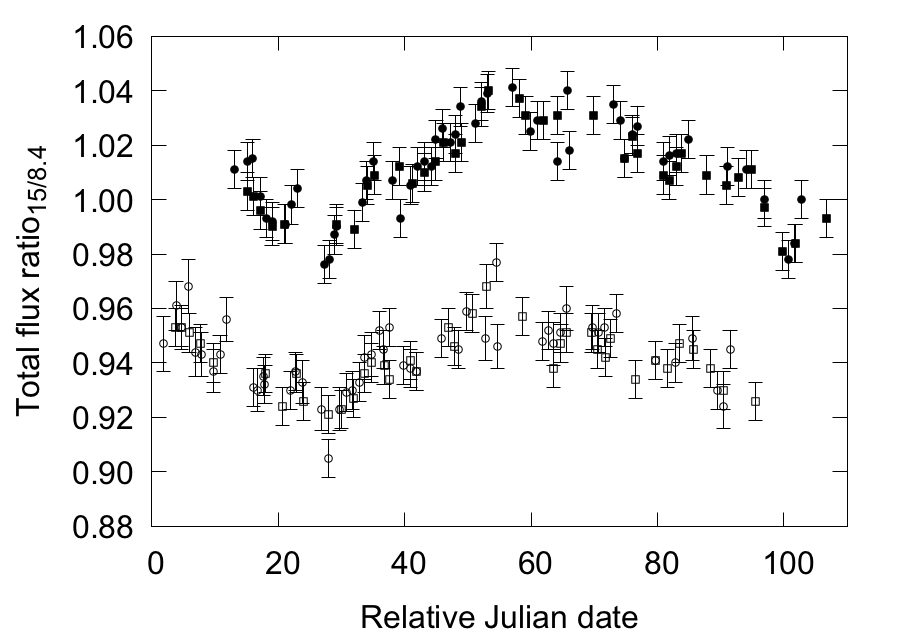}
    \includegraphics[scale=0.27]{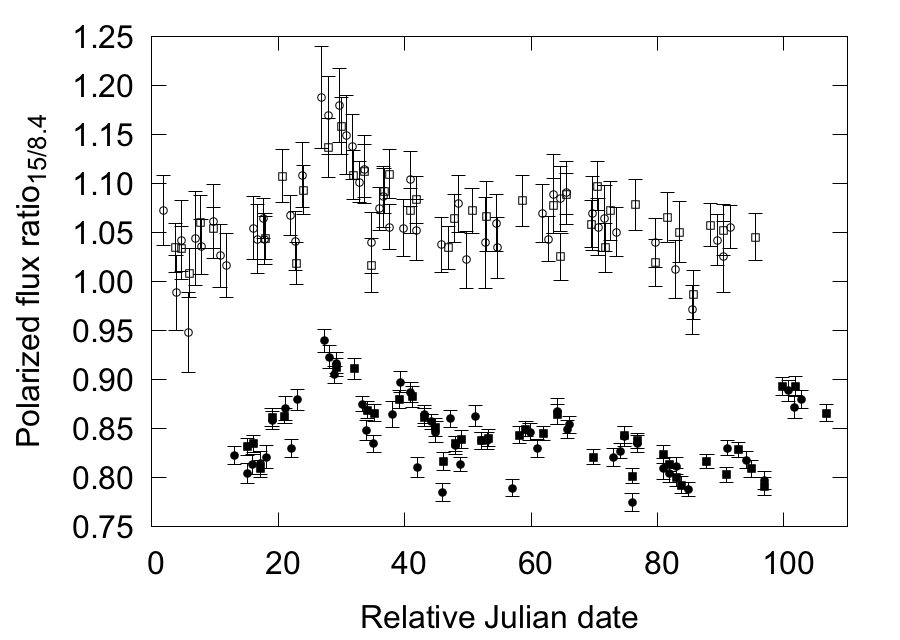}
    \includegraphics[scale=0.27]{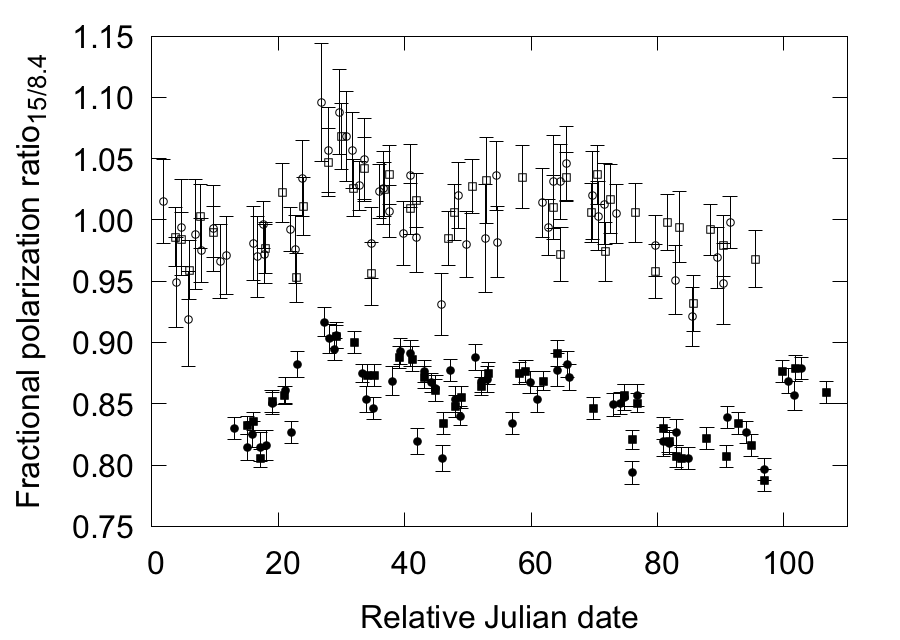}
    \includegraphics[scale=0.27]{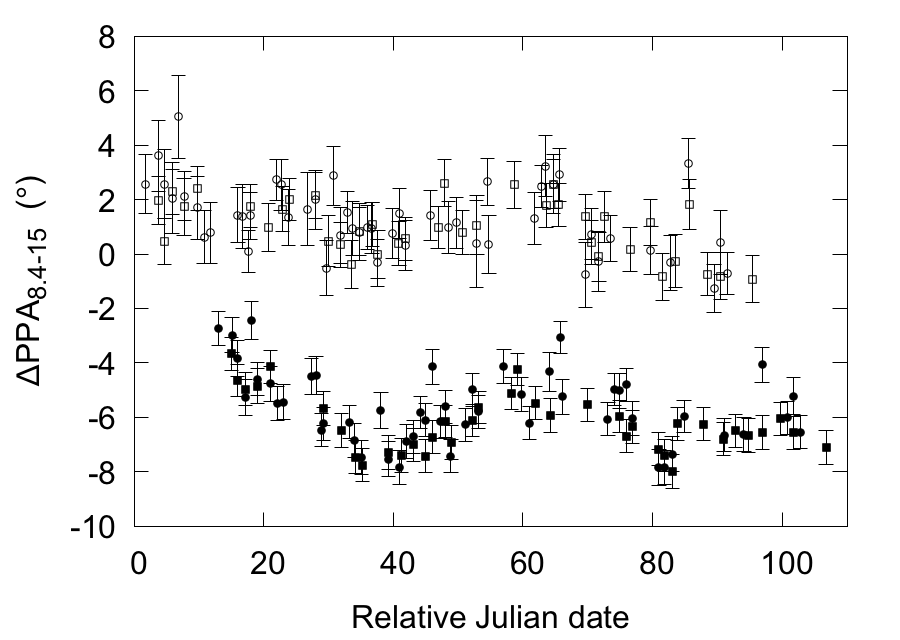}
    \caption{Flux and polarization ratios (15 / 8.4~GHz) and PPA difference ($8.4 - 15$~GHz) after removal of the time delay (11.3~d) -- image A (filled symbols), image B (unfilled), AB809 (squares), AH593 (circles). The polarized flux and fractional polarization of image A has been multiplied by 1.5 to remove most of the depolarization and make it easier to compare A and B. The PPA of image B has been rotated by 30\degr\ for a similar reason. The time delay is clearly evident for all except PPA where each image seems to vary independently. This may be a result of source variability causing different parts of the Faraday screen in the ISM of the lensing galaxy to be illuminated.}
    \label{fig:fratio}
  \end{center}
\end{figure*}

In Fig.~\ref{fig:fratio} we show that it is also possible to determine the time delay by combining the 15- and 8.4-GHz data. Correlated variations separated by the time delay are clearly seen in the ratios of total flux density, polarized flux density and percentage polarization and the CSM and CCF algorithms find $11.3 \le \tau \le 11.5$~d (Table~\ref{tab:delay}). These delays are not independent of the single-frequency values and the lower magnitude of the variations would result in larger uncertainties, but it is none-the-less reassuring that the flux-ratio datasets are consistent with our final delay estimate. The $\Delta$PPA data are discussed below.

\begin{figure*}
  \begin{center}
    \includegraphics[scale=0.24]{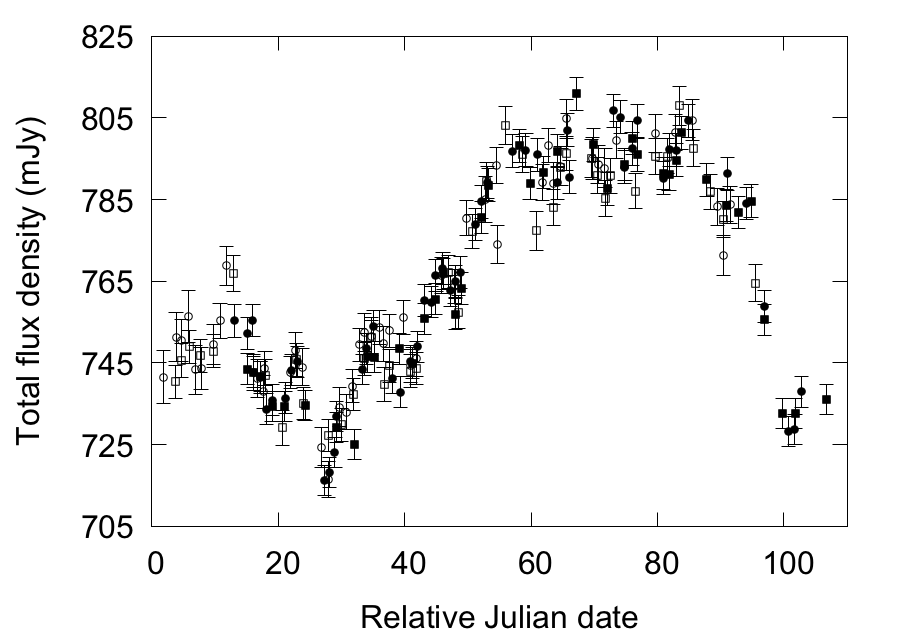}
    \includegraphics[scale=0.24]{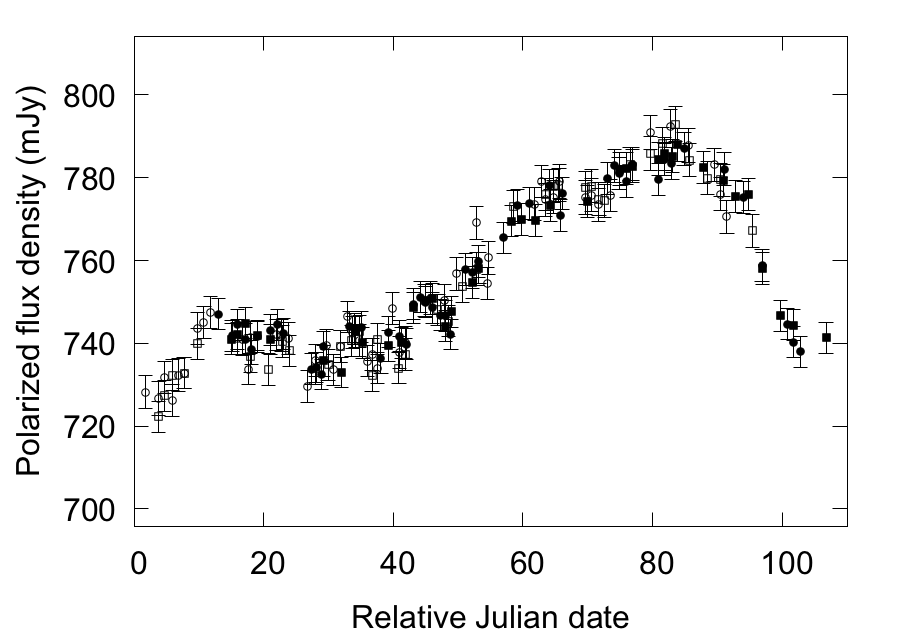}
    \includegraphics[scale=0.24]{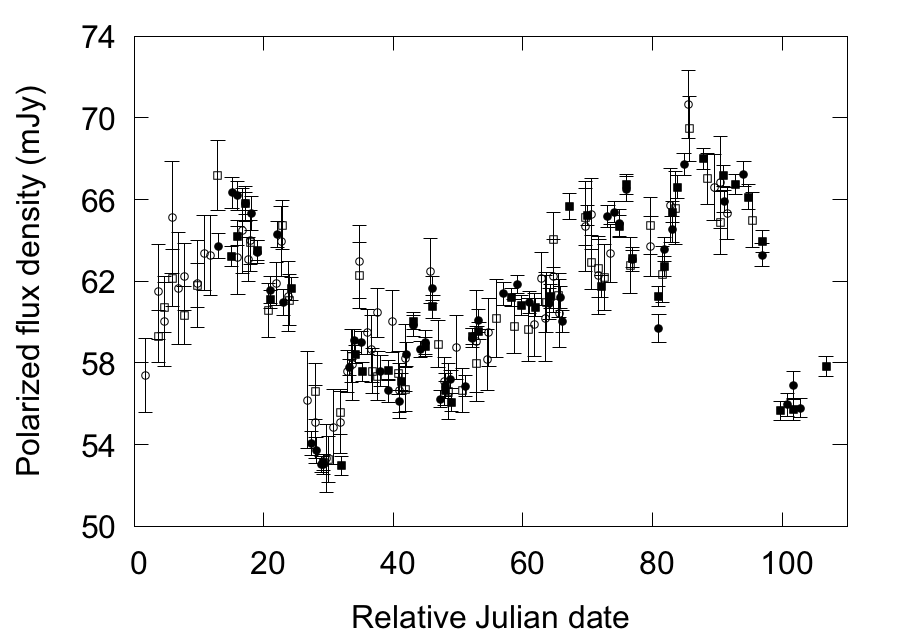}
    \includegraphics[scale=0.24]{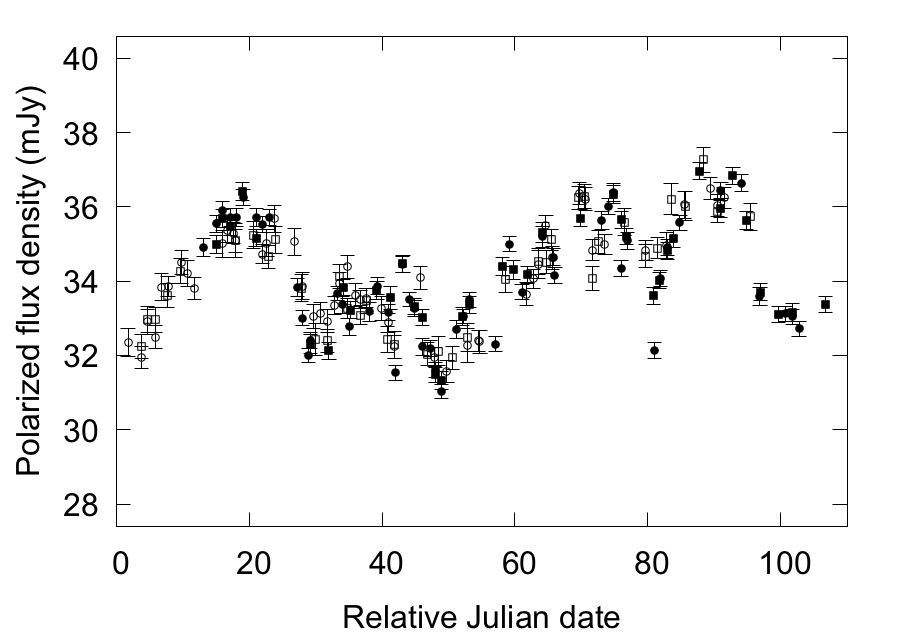}
    \includegraphics[scale=0.24]{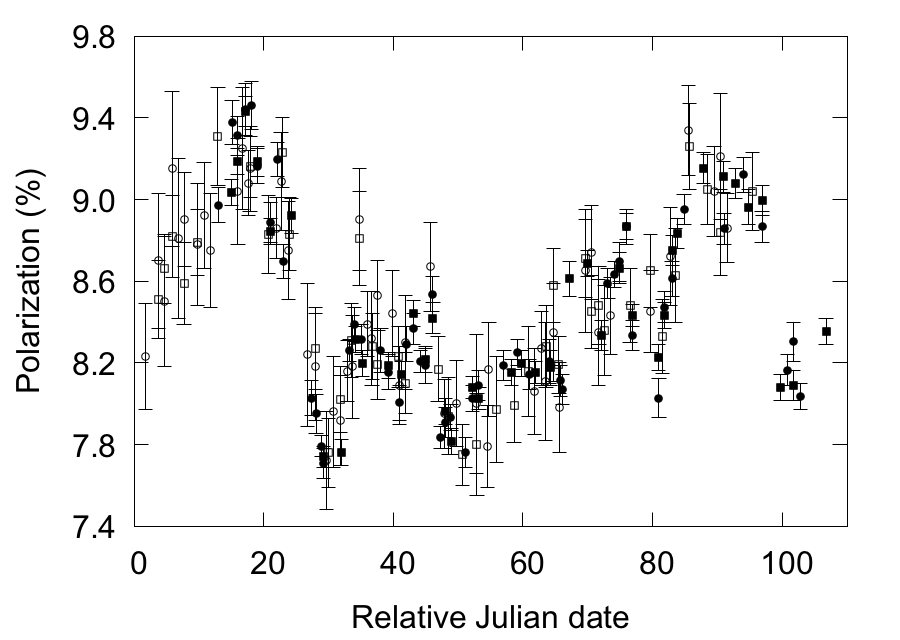}
    \includegraphics[scale=0.24]{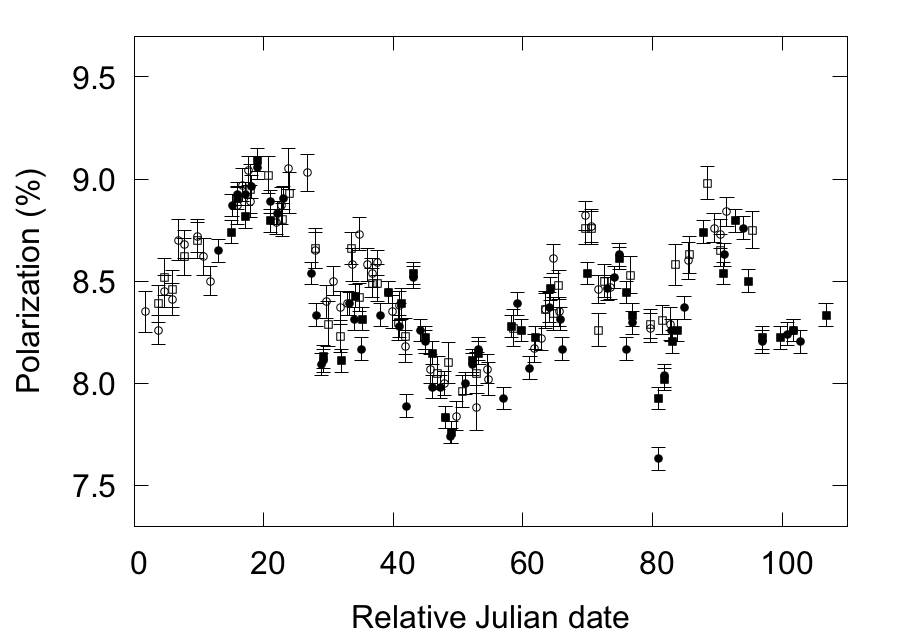}
    \includegraphics[scale=0.24]{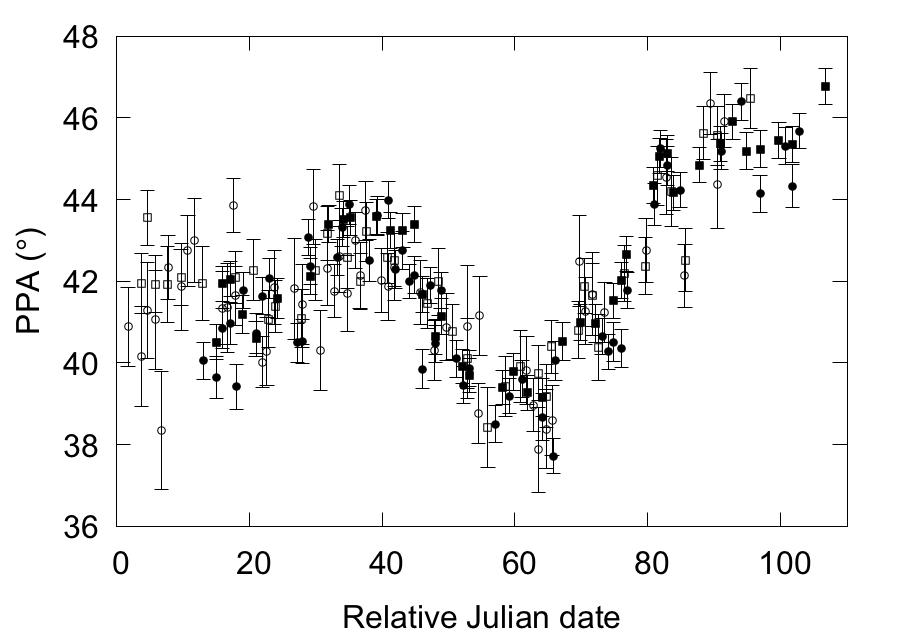}
    \includegraphics[scale=0.24]{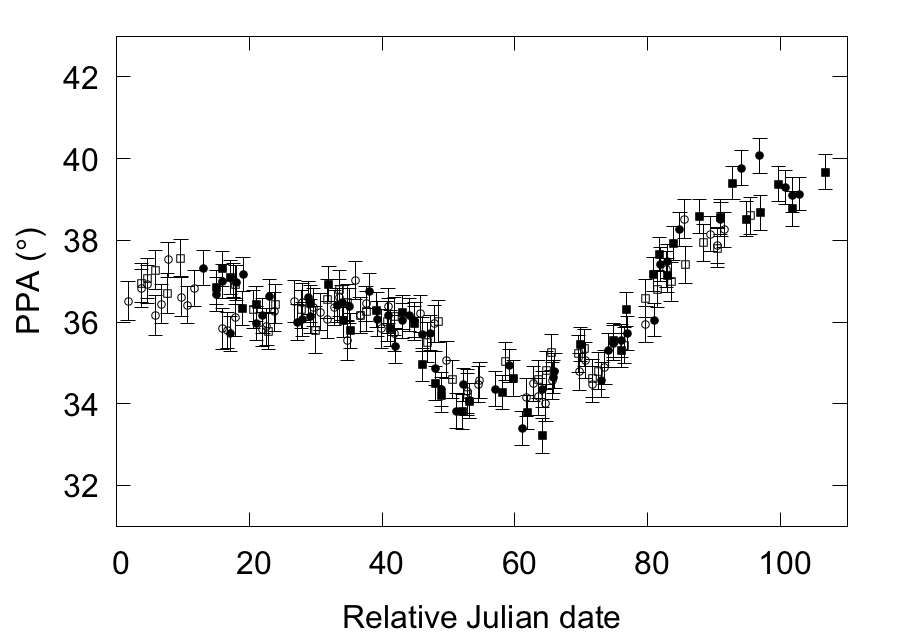}
    \caption{Combined 15-GHz (left) and 8.4-GHz (right) variability curves after removal of the time delay (11.3~d) and flux ratio/depolarization/Faraday rotation -- image A (filled symbols), image B (unfilled), AB809 (squares), AH593 (circles). From top to bottom: total flux density, polarized flux density, percentage polarization and polarization position angle. Both frequencies are plotted using the same fractional range which illustrates the larger magnitude of the variations at 15~GHz.}
    \label{fig:abahlc_comb}
  \end{center}
\end{figure*}

Fig.~\ref{fig:abahlc_comb} shows the 8.4- and 15-GHz variability curves from the 1996/1997 campaigns after the time delay and $y$-axis offset have been removed. For the latter, the image B points have been scaled to the image-A values e.g.\ the total flux density of image B has been multiplied by the A/B flux ratio. Each frequency pair (total flux density, polarized flux density, percentage polarization and PPA) has been plotted on the same vertical scale in order to highlight the increased variability at 15~GHz. The rotations applied to the PPA of image B were $-$15.5\degr\ and $-$52.5\degr\ at 15 and 8.4~GHz respectively and are the values found using the CSM technique.

The uncertainty on our time-delay estimate (0.2~d) is comparable with that of \citet{cheung14} despite their monitoring period being longer (the flaring activity covers a period of approximately 130~d) and sampling more often (every six hours). However, the $\gamma$-ray analysis is hindered in two respects relative to the radio. Firstly, \textit{Fermi} is unable to spatially resolve the two quasar images and thus each feature in the A time series is observed together with its delayed counterpart in B. This will have the effect of broadening the autocorrelation function and therefore of increasing the uncertainty in the delay. Secondly, as demonstrated in fig.~4 of \citet{cheung14}, the flux ratio in $\gamma$-rays is time variable, possibly due to microlensing \citep{sitarek16,vovk16}. Whatever the origin, it again causes the features in the $\mathrm{A}+\mathrm{B}$ time series to match more poorly than would otherwise be the case and make it more difficult to measure the time delay.

\subsection{Relative location of the radio and $\gamma$-ray emitters}

One consequence of our reanalysis is that there is no longer a significant offset between the radio ($11.3 \pm 0.2$~d) and $\gamma$-ray time delays ($11.46 \pm 0.16$~d). As noted by both \citet{cheung14} and \citet{barnacka16}, the offset implied a substantial separation (50--70~pc) between the radio and $\gamma$-ray emitting regions, but our new analysis removes the evidence for any spatial separation.

Although the location of the $\gamma$-ray emission in blazars is somewhat controversial, the similarity of the radio and $\gamma$-ray delays in B0218+357 is consistent with studies of the lags between radio, millimetre and $\gamma$-ray flares in other powerful radio sources \citep*{fuhrmann14,maxmoerbeck14,pushkarev10}. These find that although the the $\gamma$-ray emission is produced upstream from the radio emission, the offset is much smaller than that implied by the previous time-delay offset in B0218+357.

\subsection{The Faraday rotation measure of A and B}

The true rotation measure of images A and B has been uncertain since the discovery that B0218+357 is a gravitational lens. \citet{odea92} found that the PPA of each image rotated in opposite directions, but included data from \citet{patnaik93} which were known to contain systematic errors (thus preventing those authors from measuring the RM of each image individually) and which were observed at a completely different epoch ($\Delta t \sim 1.5$~yr). A later paper \citep{patnaik01} claimed much higher values of $\ga$8000~rad\,m$^{-2}$ for each image.

\begin{figure}
  \begin{center}
    \includegraphics[scale=0.35]{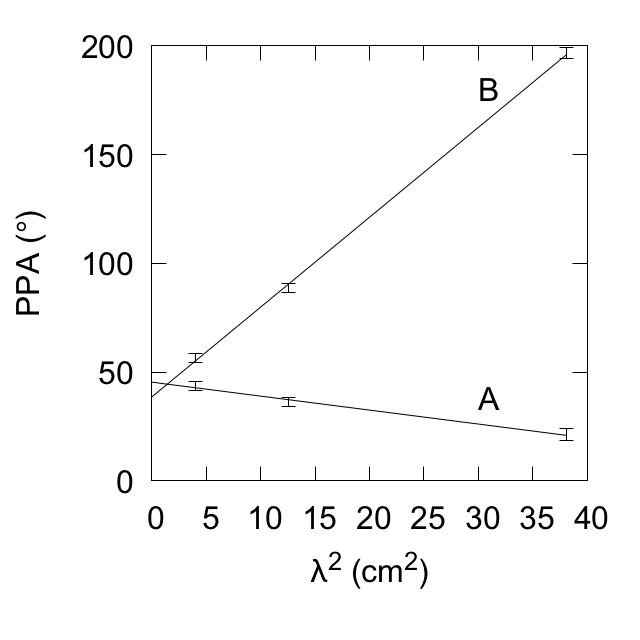}
    \caption{PPA versus $\lambda^2$ of images A and B for the epoch observed on 5 November as part of the AB809 monitoring. An absolute calibration error of 2\degr\ has been added in quadrature to each measurement. The linear fits correspond to rotation measures of $\mathrm{RM_A} = -112 \pm 11$ and $\mathrm{RM_B} = 720 \pm 18$~rad\,m$^{-2}$.}
    \label{fig:rm}
  \end{center}
\end{figure}

Although the PPA monitoring data presented here were not included in the time-delay analysis, they still remain useful for investigating the RM of each image. The PPA is plotted in Fig.~\ref{fig:rm} as a function of $\lambda^2$ for a single representative epoch (5 November) of the AB809 data. The variation of the PPA with $\lambda^2$ is clearly linear and fits to the data give $\mathrm{RM_A} = -112 \pm 11$ and $\mathrm{RM_B} = 720 \pm 18$~rad\,m$^{-2}$. Our own analysis of the \citet{patnaik01} data (VLA project code AP366) gives values within 1--2~$\sigma$ of our own, $\mathrm{RM_A} = -166 \pm 20$ and $\mathrm{RM_B} = 693 \pm 11$~rad\,m$^{-2}$, and we suspect that \citeauthor{patnaik01} inadvertently swapped the $Q$ and $U$ images for one of the frequencies.

We therefore agree with \citet{odea92} that the PPAs of each image rotate in opposite directions, but find that the RM of image B is many times larger than that of A. The contribution from our own galaxy along the direction to 0218+357 is estimated to be $-64$~rad\,m$^{-2}$ \citep*{taylor09} and thus the intrinsic RM ratio is even larger. Although absorbing gas is only seen in front of image A, there must be more ionizing gas in front of image B, a stronger or more ordered magnetic field, or some combination of these. The different sense of rotation means that the components of the magnetic field perpendicular to the line of sight are antiparallel to each other at the locations of A and B. As the lensing galaxy is a nearly face-on spiral, polarization observations probe the magnetic field perpendicular to the galactic disc.

The $\Delta$PPA plot in Fig.~\ref{fig:fratio} suggests that independent RM variations are present in one or both images, possibly due to motion in the lensed source causing different parts of the Faraday screen to be illuminated. However, the error bars are quite large and we remain unconvinced that the effect is real.

\subsection{Implications for $H_0$}

Determining $H_0$ from 0218+357 was long hampered by the uncertain position of the lensing galaxy, this being difficult to measure due to the very small separation of the lensed images which therefore obscure the core of the galaxy \citep*{jackson00,lehar00}. A subsequent attempt was made to directly measure the lens position using the Advanced Camera for Surveys onboard the \textit{Hubble Space Telescope} \citep{york05} but the results depend on whether the spiral arms are masked or not. The corresponding value of $H_0$ changes by $\ga$10~per~cent depending on the approach used. \citet{cheung14} combined their $\gamma$-ray delay with one of the six lens models presented by \citeauthor{york05} (no masking, inclusion of additional constraints from the VLBI substructure and a non-isothermal mass distribution).

Although not a direct method of determining the lens centre, an alternative approach involves using the complete Einstein ring as a source of model constraints for the \textsc{LensClean} algorithm \citep{kochanek92,wucknitz04a}. This gives more constraints than traditional modelling where only the relative positions and flux ratios of the cores (including the VLBI substructure) are used. The ring also samples the gravitational potential at all position angles relative to the centre of the lensing galaxy. We therefore consider this to be the most reliable lens model published to date. Using a VLA 15-GHz image and assuming that the lens is isothermal, \citet{wucknitz04b} were able to measure the position of the lensing galaxy with a 1-$\sigma$ accuracy of $\sim$5~mas which, when combined with our new time delay, results in $H_0 = 72.9 \pm 2.6$~km\,s$^{-1}$\,Mpc$^{-1}$. As shown by \citeauthor{wucknitz04b}, the value of $H_0$ for B0218+357 found using LensClean is not expected to change very much even if a non-isothermal model is used.

This value is very close to that found locally using Cepheid-calibrated supernovae \citep[$73.24 \pm 1.74$~km\,s$^{-1}$\,Mpc$^{-1}$ --][]{riess16} and is compatible at 1~$\sigma$ with the most recent \textit{Planck} result \citep[$67.8 \pm 0.9$~km\,s$^{-1}$\,Mpc$^{-1}$ --][]{planck16}. The huge progress in reducing the uncertainties of the local and Cosmic Microwave Background-based values has led to formal disagreement between the two, the origin of which is unclear. The relatively large error on the 0218+357 value leaves it compatible with both, but it should be possible to reduce the uncertainty further by applying the \textsc{LensClean} algorithm to higher-resolution VLA 15-GHz data that include the Pie Town antenna. The final analysis of these data has not yet been completed (O.~Wucknitz, private communication).

\section{Conclusions}
\label{conclusions}

We have performed a reanalysis of the VLA monitoring data for the gravitational lens JVAS~B0218+357. Combining the two monitoring campaigns observed concurrently over a 100-d period from October 1996 to January 1997 has allowed us to measure a time delay of $11.3 \pm 0.2$~d (1~$\sigma$). This is significantly higher than the radio measurements published using the data from each campaign separately and is consistent with the recently measured $\gamma$-ray value. One consequence of this is there is no longer any evidence of an offset between the radio and $\gamma$-ray emitting regions \citep{cheung14,barnacka16}. When combined with the lens model of \citet{wucknitz04b}, the new time delay gives $H_0 = 72.9 \pm 2.6$~km\,s$^{-1}$\,Mpc$^{-1}$ (1~$\sigma$).

One of the biggest changes compared to the original B99 analysis is accurate modelling and subtraction of the Einstein ring emission and the subsequent improvement in the polarization variability curves. Even at 5~GHz it has been possible to obtain useful polarization measurements. Having three frequencies with good absolute-PPA calibration has also allowed us to calculate the rotation measures of each image accurately. The PPAs of each image rotate in opposite directions and the RM at the position of image B is several times larger than that of A. Much more could be learned about the magnetic field of the $z = 0.6847$ spiral lensing galaxy by mapping the rotation measure throughout the Einstein ring. This is a challenging observation due to the faintness and small diameter of the ring, but e-MERLIN operating in its 4--8-GHz band probably offers the best combination of sensitivity and angular resolution.

Two sources of systematic error have been uncovered -- the PPA measurements are offset as a function of hour angle and proximity of the source to the zenith and the D-term calibration can also produce PPA offsets, although the latter were only seen with the C00 data at 8.4~GHz. Although both can be corrected for, the origin of both effects is still not fully understood and until they are will complicate analysis and interpretation of future polarization monitoring observations. This is particularly the case for B0218+357 for which convenient calibrators naturally lie close to the zenith, but we caution that the popular `unpolarized' VLA calibrators, 3C~48 and OQ~208, both pass within 10\degr\ of the zenith.

\section*{Acknowledgements}

The authors would like to thank Robert Laing for many discussions that helped enormously in the preparation of this paper, Rick Perley for his advice on the VLA's polarization characteristics and Olaf Wucknitz for his thoughts regarding the lens model. The National Radio Astronomy Observatory is a facility of the National Science Foundation operated under cooperative agreement by Associated Universities, Inc. This research has made use of data from the University of Michigan Radio Astronomy Observatory which has been supported by the University of Michigan and by a series of grants from the National Science Foundation, most recently AST-0607523.




\bibliographystyle{mnras}
\bibliography{lensing}

\begin{thebibliography}{}

\bibitem[\protect\citeauthoryear{{Barnacka} et~al.}{{Barnacka}
  et~al.}{2016}]{barnacka16}
{Barnacka} A., {Geller} M.~J., {Dell'Antonio} I.~P.,  {Zitrin} A., 2016, \apj,
  821, 58

\bibitem[\protect\citeauthoryear{{Biggs} et~al.}{{Biggs}
  et~al.}{1999}]{biggs99}
{Biggs} A.~D., {Browne} I.~W.~A., {Helbig} P., {Koopmans} L.~V.~E., {Wilkinson}
  P.~N.,  {Perley} R.~A., 1999, \mnras, 304, 349

\bibitem[\protect\citeauthoryear{{Bignell}}{{Bignell}}{1982}]{bignell82}
{Bignell} C., 1982, in {Thompson} A.~R.,  {D'Addario} L.~R., ed, Synthesis
  Mapping

\bibitem[\protect\citeauthoryear{{Browne} et~al.}{{Browne}
  et~al.}{1993}]{browne93}
{Browne} I.~W.~A., {Patnaik} A.~R., {Walsh} D.,  {Wilkinson} P.~N., 1993,
  \mnras, 263, L32

\bibitem[\protect\citeauthoryear{{Cheung} et~al.}{{Cheung}
  et~al.}{2014}]{cheung14}
{Cheung} C.~C. et~al., 2014, \apjl, 782, L14

\bibitem[\protect\citeauthoryear{{Cohen} et~al.}{{Cohen}
  et~al.}{2000}]{cohen00}
{Cohen} A.~S., {Hewitt} J.~N., {Moore} C.~B.,  {Haarsma} D.~B., 2000, \apj,
  545, 578

\bibitem[\protect\citeauthoryear{{Cohen}, {Lawrence} \& {Blandford}}{{Cohen}
  et~al.}{2003}]{cohen03}
{Cohen} J.~G., {Lawrence} C.~R.,  {Blandford} R.~D., 2003, \apj, 583, 67

\bibitem[\protect\citeauthoryear{{Corbett} et~al.}{{Corbett}
  et~al.}{1996}]{corbett96}
{Corbett} E.~A., {Browne} I.~W.~A., {Wilkinson} P.~N.,  {Patnaik} A., 1996, in
  IAU Symposium, Vol. 173, {Kochanek} C.~S.,  {Hewitt} J.~N., ed, Astrophysical
  Applications of Gravitational Lensing, p.~37

\bibitem[\protect\citeauthoryear{{Fuhrmann} et~al.}{{Fuhrmann}
  et~al.}{2014}]{fuhrmann14}
{Fuhrmann} L. et~al., 2014, \mnras, 441, 1899

\bibitem[\protect\citeauthoryear{{Jackson}, {Xanthopoulos} \&
  {Browne}}{{Jackson} et~al.}{2000}]{jackson00}
{Jackson} N., {Xanthopoulos} E.,  {Browne} I.~W.~A., 2000, \mnras, 311, 389

\bibitem[\protect\citeauthoryear{{King} et~al.}{{King} et~al.}{1999}]{king99}
{King} L.~J., {Browne} I.~W.~A., {Marlow} D.~R., {Patnaik} A.~R.,  {Wilkinson}
  P.~N., 1999, \mnras, 307, 225

\bibitem[\protect\citeauthoryear{{Kochanek} \& {Narayan}}{{Kochanek} \&
  {Narayan}}{1992}]{kochanek92}
{Kochanek} C.~S.,  {Narayan} R., 1992, \apj, 401, 461

\bibitem[\protect\citeauthoryear{{Kundi{\'c}} et~al.}{{Kundi{\'c}}
  et~al.}{1997}]{kundic97}
{Kundi{\'c}} T. et~al., 1997, \apj, 482, 75

\bibitem[\protect\citeauthoryear{{Leh{\'a}r} et~al.}{{Leh{\'a}r}
  et~al.}{2000}]{lehar00}
{Leh{\'a}r} J. et~al., 2000, \apj, 536, 584

\bibitem[\protect\citeauthoryear{{Mantovani} et~al.}{{Mantovani}
  et~al.}{2010}]{mantovani10}
{Mantovani} F., {Rossetti} A., {Junor} W., {Saikia} D.~J.,  {Salter} C.~J.,
  2010, \aap, 518, A33

\bibitem[\protect\citeauthoryear{{Max-Moerbeck} et~al.}{{Max-Moerbeck}
  et~al.}{2014}]{maxmoerbeck14}
{Max-Moerbeck} W. et~al., 2014, \mnras, 445, 428

\bibitem[\protect\citeauthoryear{{Mittal} et~al.}{{Mittal}
  et~al.}{2006}]{mittal06}
{Mittal} R., {Porcas} R., {Wucknitz} O., {Biggs} A.,  {Browne} I., 2006, \aap,
  447, 515

\bibitem[\protect\citeauthoryear{{Nan} et~al.}{{Nan} et~al.}{1999}]{nan99}
{Nan} R., {Gabuzda} D.~C., {Kameno} S., {Schilizzi} R.~T.,  {Inoue} M., 1999,
  \aap, 344, 402

\bibitem[\protect\citeauthoryear{{Nan Ren-Dong} et~al.}{{Nan Ren-Dong}
  et~al.}{1991}]{nan91}
{Nan Ren-Dong} , {Schilizzi} R.~T., {van Breugel} W.~J.~M., {Fanti} C., {Fanti}
  R., {Muxlow} T.~W.~B.,  {Spencer} R.~E., 1991, \aap, 245, 449

\bibitem[\protect\citeauthoryear{{O'Dea} et~al.}{{O'Dea} et~al.}{1992}]{odea92}
{O'Dea} C.~P., {Baum} S.~A., {Stanghellini} C., {Dey} A., {van Breugel} W.,
  {Deustua} S.,  {Smith} E.~P., 1992, \aj, 104, 1320

\bibitem[\protect\citeauthoryear{{Patnaik} et~al.}{{Patnaik}
  et~al.}{1993}]{patnaik93}
{Patnaik} A.~R., {Browne} I.~W.~A., {King} L.~J., {Muxlow} T.~W.~B., {Walsh}
  D.,  {Wilkinson} P.~N., 1993, \mnras, 261, 435

\bibitem[\protect\citeauthoryear{{Patnaik} et~al.}{{Patnaik}
  et~al.}{1992}]{patnaik92}
{Patnaik} A.~R., {Browne} I.~W.~A., {Wilkinson} P.~N.,  {Wrobel} J.~M., 1992,
  \mnras, 254, 655

\bibitem[\protect\citeauthoryear{{Patnaik} et~al.}{{Patnaik}
  et~al.}{2001}]{patnaik01}
{Patnaik} A.~R., {Menten} K.~M., {Porcas} R.~W.,  {Kemball} A.~J., 2001, in
  Astronomical Society of the Pacific Conference Series, Vol. 237, {Brainerd}
  T.~G.,  {Kochanek} C.~S., ed, Gravitational Lensing: Recent Progress and
  Future Goals, p.~99

\bibitem[\protect\citeauthoryear{{Pelt} et~al.}{{Pelt} et~al.}{1994}]{pelt94}
{Pelt} J., {Hoff} W., {Kayser} R., {Refsdal} S.,  {Schramm} T., 1994, \aap,
  286, 775

\bibitem[\protect\citeauthoryear{{Pelt} et~al.}{{Pelt} et~al.}{1996}]{pelt96}
{Pelt} J., {Kayser} R., {Refsdal} S.,  {Schramm} T., 1996, \aap, 305, 97

\bibitem[\protect\citeauthoryear{{Perley} \& {Butler}}{{Perley} \&
  {Butler}}{2013a}]{perley13a}
{Perley} R.~A.,  {Butler} B.~J., 2013a, \apjs, 204, 19

\bibitem[\protect\citeauthoryear{{Perley} \& {Butler}}{{Perley} \&
  {Butler}}{2013b}]{perley13b}
{Perley} R.~A.,  {Butler} B.~J., 2013b, \apjs, 206, 16

\bibitem[\protect\citeauthoryear{{Peterson} et~al.}{{Peterson}
  et~al.}{1998}]{peterson98}
{Peterson} B.~M., {Wanders} I., {Horne} K., {Collier} S., {Alexander} T.,
  {Kaspi} S.,  {Maoz} D., 1998, \pasp, 110, 660

\bibitem[\protect\citeauthoryear{{Planck Collaboration} et~al.}{{Planck
  Collaboration} et~al.}{2016}]{planck16}
{Planck Collaboration}  et~al., 2016, \aap, 594, A13

\bibitem[\protect\citeauthoryear{{Pushkarev}, {Kovalev} \&
  {Lister}}{{Pushkarev} et~al.}{2010}]{pushkarev10}
{Pushkarev} A.~B., {Kovalev} Y.~Y.,  {Lister} M.~L., 2010, \apjl, 722, L7

\bibitem[\protect\citeauthoryear{{Riess} et~al.}{{Riess}
  et~al.}{2016}]{riess16}
{Riess} A.~G. et~al., 2016, \apj, 826, 56

\bibitem[\protect\citeauthoryear{{Shepherd}}{{Shepherd}}{1997}]{shepherd97}
{Shepherd} M.~C., 1997, in Astronomical Society of the Pacific Conference
  Series, Vol. 125, {Hunt} G.,  {Payne} H., ed, Astronomical Data Analysis
  Software and Systems VI, p.~77

\bibitem[\protect\citeauthoryear{{Sitarek} \& {Bednarek}}{{Sitarek} \&
  {Bednarek}}{2016}]{sitarek16}
{Sitarek} J.,  {Bednarek} W., 2016, \mnras, 459, 1959

\bibitem[\protect\citeauthoryear{{Taylor}, {Stil} \& {Sunstrum}}{{Taylor}
  et~al.}{2009}]{taylor09}
{Taylor} A.~R., {Stil} J.~M.,  {Sunstrum} C., 2009, \apj, 702, 1230

\bibitem[\protect\citeauthoryear{{Tewes}, {Courbin} \& {Meylan}}{{Tewes}
  et~al.}{2013}]{tewes13}
{Tewes} M., {Courbin} F.,  {Meylan} G., 2013, \aap, 553, A120

\bibitem[\protect\citeauthoryear{{Vovk} \& {Neronov}}{{Vovk} \&
  {Neronov}}{2016}]{vovk16}
{Vovk} I.,  {Neronov} A., 2016, \aap, 586, A150

\bibitem[\protect\citeauthoryear{{White} \& {Peterson}}{{White} \&
  {Peterson}}{1994}]{white94}
{White} R.~J.,  {Peterson} B.~M., 1994, \pasp, 106, 879

\bibitem[\protect\citeauthoryear{{Wucknitz}}{{Wucknitz}}{2004}]{wucknitz04a}
{Wucknitz} O., 2004, \mnras, 349, 1

\bibitem[\protect\citeauthoryear{{Wucknitz}, {Biggs} \& {Browne}}{{Wucknitz}
  et~al.}{2004}]{wucknitz04b}
{Wucknitz} O., {Biggs} A.~D.,  {Browne} I.~W.~A., 2004, \mnras, 349, 14

\bibitem[\protect\citeauthoryear{{York} et~al.}{{York} et~al.}{2005}]{york05}
{York} T., {Jackson} N., {Browne} I.~W.~A., {Wucknitz} O.,  {Skelton} J.~E.,
  2005, \mnras, 357, 124

\end{thebibliography}


\bsp	
\label{lastpage}
\end{document}